\documentclass{aa}
\usepackage{graphicx}
\usepackage{txfonts}
\usepackage{natbib} 

\bibpunct{(}{)}{;}{a}{}{,} % to follow the A&A style

\newcommand{\CtHt}{\mbox{C$_2$H$_2$}}

\newcommand{\Wo}{\mbox{$EW_{3.1}$}}
\newcommand{\Wf}{\mbox{$EW_{3.5}$}}
\newcommand{\We}{\mbox{$EW_{3.8}$}}
\newcommand{\Ws}{\mbox{$EW_{\mathrm{SiO}}$}}

\def\oversim#1#2{\lower0.5pt\vbox{\baselineskip0pt \lineskip-0.5pt
     \ialign{$\mathsurround0pt #1\hfil##\hfil$\crcr#2\crcr\sim\crcr}}}
\def\gsim{\mathrel{\mathpalette\oversim>}}    % > over \sim
\begin{document}
  \title{Three-micron spectra of AGB stars and supergiants in nearby galaxies
\thanks{Based on observations at the European Southern Observatory,
on the Very Large Telescope with an instrument, ISAAC
(projects 68.D-0660, 69.D-0623)}
}

%   \subtitle{}

\author{M.~Matsuura\inst{1}, A.A.~Zijlstra\inst{1}, J.Th.~van~Loon\inst{2},
I.~Yamamura\inst{3}, 
A.J.~Markwick\inst{4}, 
P.A.~Whitelock\inst{5}, 
P.M.~Woods\inst{1}, 
J.R.~Marshall\inst{2},
M.W.~Feast\inst{6}, 
    \and  
L.B.F.M.~Waters 
\inst{7,8}, 
          }

   \offprints{M.Matsuura}

   \institute{       
       School of Physics and Astronomy, University of Manchester,
       Sackville Street, P.O. Box 88, Manchester M60 1QD, UK \\
              \email{M.M. (m.matsuura@manchester.ac.uk)}
\and
       Astrophysics Group,
       School of Chemistry and Physics, Keele University, Staffordshire ST5 5BG,
       UK
\and
       The Institute of Space and Astronautical Science, 
       Japan Aerospace Exploration Agency,
       Yoshino-dai 3-1-1, Sagamihara, Kanagawa 229-8510, Japan
\and
       Space Science Division, NASA Ames Research Centre, MS 245-3,
       Moffett Field, CA 94035, USA
\and
       South African Astronomical Observatory, P.O.Box 9, 7935     
	Observatory, Republic of South Africa
\and
       Astronomy Department, University of Cape Town, 7701,
       Rondebosch, South Africa
\and       
	Astronomical Institute `Anton Pannekoek', University of Amsterdam,       
	Kruislaan 403, 1098 SJ, Amsterdam, the Netherlands
\and
       Instituut voor Sterrenkunde, Katholieke Universiteit Leuven, 
       Celestijnenlaan 200B, 3001 Heverlee, Belgium
             }

   \authorrunning{Matsuura et al.}
   \titlerunning{Infrared spectra of AGB stars in nearby galaxies}

\date{Received 2004; accepted}

   \abstract{The dependence of stellar molecular bands 
     on the metallicity is studied using infrared $L$-band spectra of 
     AGB stars (both carbon-rich and oxygen-rich) and M-type
     supergiants in the Large and Small Magellanic Clouds (LMC and
     SMC) and in the Sagittarius Dwarf Spheroidal Galaxy.  The
     spectra cover SiO bands for
     oxygen-rich stars, and acetylene (C$_2$H$_2$), CH and HCN bands for
     carbon-rich AGB stars.  
     The equivalent width of acetylene is found to be high even at low
     metallicity.  The high C$_2$H$_2$ abundance can be explained with
     a high carbon-to-oxygen (C/O) ratio for lower metallicity carbon
     stars.  In contrast, the HCN equivalent width is low: fewer than half
     of the extra-galactic carbon stars show the 3.5\,$\mu$m HCN band,
     and only a few LMC stars show high HCN equivalent width.  HCN
     abundances are limited by both nitrogen and carbon elemental
     abundances. The amount of synthesized nitrogen depends on the
     initial mass, and stars with high luminosity (i.e. high initial 
     mass) could have a high HCN abundance. CH bands are found in both
     the extra-galactic and Galactic carbon stars.
     One SMC post-AGB star, SMC-S2, shows the 3.3\,$\mu$m PAH band.
     This first detection of a PAH band from an SMC post-AGB
     star confirms PAHs can form in these low-metallicity stars.
     None of the oxygen-rich LMC stars show SiO bands, except one
     possible detection in a low quality spectrum.  The limits on the
     equivalent widths of the SiO bands are below the     
     expectation of up to 30\,\AA\, for LMC metallicity.  Several
     possible explanations are discussed, mostly based on the effect of
     pulsation and circumstellar dust.
     The observations imply that LMC and
     SMC carbon stars could reach mass-loss rates as high as their
     Galactic counterparts,  because  there are more carbon atoms available
     and more carbonaceous dust can be formed. On the other hand,
     the lack of SiO suggests less dust and lower mass-loss rates in
     low-metallicity oxygen-rich stars. The effect on the
     ISM dust enrichment is discussed.
   }

%
%________________________________________________________________

\maketitle

%\tableofcontents

\section{ Introduction}

Asymptotic Giant 
Branch (AGB) and Red Supergiant (RSG) stars are
important sources of dust 
and gas in galaxies.  Within the final
$10^4$--$10^5$ years of the AGB 
phase, stars lose up to 80\,\% of
their mass via a slow wind with mass-loss 
rates up to $10^{-4}\, \rm
M_\odot\,yr^{-1}$.  Red supergiants can 
experience a similar mass-loss
phase. AGB stars have initial masses of 
1--8\,$M_{\sun}$, but their large
population compensates for the lower 
mass.  
The other main source of dust and gas, supernovae, 
produce higher elements per event
but occur at a much lower rate. The balance 
between dust production by
supernovae and by AGB/RSG stars is not well 
known. If the mass-loss rate
of AGB/RSG stars is much lower at 
low-metallicity \citep{Zijlstra04a},
the relative contributions may be 
strongly dependent on metallicity
\citep{Edmunds2001}.  

How  mass-loss 
rates are affected by metallicity is not well
understood.  In general, the elemental abundances of extra-galactic
stars are assumed to scale with the metallicity, but this
ignores the elements which are synthesized in the AGB stars or
supergiants themselves. The wind is dust-driven, and dust forms out of
simple molecules that are stable at the photospheric temperatures of the 
cool giants.  An indication of the metallicity effect on mass loss can 
therefore be obtained by studying the molecules in AGB and RSG
stars at low metallicity. 

AGB stars are classified into two major spectral types: carbon-rich
stars and oxygen-rich stars.  In photospheric chemistry, CO
molecules are formed first as their bonding energy is highest.  In a
carbon-rich environment, carbon atoms remaining after CO formation are 
incorporated into carbon-bearing molecules, such as C$_2$, C$_2$H$_2$, CN, 
and HCN. Oxygen-bearing molecules such as H$_2$O, SiO, CO$_2$ are formed in 
oxygen-rich stars.  There is additionally a minor group of stars,
S-type stars, which 
have carbon-to-oxygen ratio (C/O ratio) close to unity, and where CO molecules are 
prominent in the infrared spectra.

Previous surveys of AGB stars in the local group galaxies have shown
that the ratio of carbon-rich stars to oxygen-rich ones depends on the
metallicity of the galaxies \citep[e.g.][]{Groenewegen99}.  In the
solar neighbourhood, about 90\,\% of the late AGB stars are oxygen-rich,
while in the SMC, which has lower metallicity, about 80\,\% are carbon-rich.  This is because a
fixed amount of carbon injected into the envelope has a larger effect
for a low metallicity, leading to a higher C/O ratio
\citep{Lattanzio03}, but also because dredge-up of newly synthesized
carbon atoms is more efficient in AGB stars of lower metallicity
\citep{Vassiliadis93}.  \citet{Marigo02} concludes that the
carbon-to-oxygen (C/O) ratio should be systematically higher in
carbon-rich stars at lower metallicity.  Spectroscopic evidence of high 
C/O ratio could be found in data of extra-galactic AGB stars.
\citet*{vanLoon99a} presented the first L-band spectra of LMC carbon
stars, and found that the equivalent widths of the 3.1\,$\mu$m HCN and
C$_2$H$_2$ bands appear to be similar in LMC stars and in Galactic
stars.  \citet[][hereafter Paper I]{Matsuura02b} found that the
equivalent widths of the 3.1\,$\mu$m band and the 3.8\,$\mu$m C$_2$H$_2$ 
band are larger in LMC stars than in Galactic stars.  This shows that 
abundances of carbon-bearing molecules do not scale simply with the 
metallicity of the host galaxies; instead, carbon synthesized inside AGB 
stars should result in the higher C/O ratio in the LMC stars.  The 
high C/O ratio compensates for the low elemental abundances: carbon-bearing 
molecules are abundant in carbon stars at
lower metallicity.

 Here we present an observational study of infrared molecular bands in
AGB/RSG stars in nearby galaxies.  The target galaxies are the Large
Magellanic Cloud (LMC) and the Small Magellanic Cloud (SMC), with
metallicities of about half and one-quarter of the solar value,
respectively, and the Sagittarius Dwarf Spheroidal Galaxy (SgrD), where
the two planetary nebulae and the AGB stars have [Fe/H]$=-0.55$
\citep{Dudziak00}; colour-magnitude diagrams show a range of
metallicities down to [Fe/H]\,=\,$-1.15$ \citep{Marconi98}.  The
spectra target HCN and C$_2$H$_2$ bands from carbon-rich stars and SiO 
bands from oxygen-rich stars.  We choose these molecules because they are 
parent molecules for dust grain formation (via PAH and silicate), and their 
abundance could affect the dust formation efficiency and, therefore, the 
mass-loss rates of the dust-driven winds of these stars.

\section{ 
Observations}

We selected AGB stars and late-type supergiants in the LMC, the SMC
and the SgrD from published catalogues: \citet{Trams99b, vanLoon99a} for
LMC stars, \citet{Groenewegen98} for SMC stars, and \citet{Whitelock96,
Whitelock99} for SgrD stars.  The selections are based on L-band magnitudes
and known classification (oxygen-rich or carbon-rich).  The observed
targets are listed in Table\,\ref{table-target}.  The LMC and the SMC
contain luminous and high mass-loss stars that are likely to be young
(intermediate-mass) stars. The SgrD stars are older, with a likely age
around 5\,Gyr, and are optical carbon stars without evidence for high
mass loss.

L-band spectra were observed with ISAAC on the Very Large Telescope
(VLT) at ESO Paranal, Chile on 12th--14th of December 2001 (LMC
targets) and 24th of July 2002 (SMC and SgrD targets).  The sky
conditions were clear on December 13th, and thin cloudy on the 12th and
14th of December and on the 24th July.  On cloudy nights, especially on July
24th, the sky conditions were variable, and cancellation of atmospheric 
telluric lines was not ideal.  We used chop-and-nod to
subtract the sky 
background.  Jitter is used along the slit to
minimize the influence of hot 
pixels.

We used two ISAAC instrument modes: low spectral resolution (LR) and
medium resolution (MR).  The LR mode covers the entire L-band, so this 
mode was used mainly for carbon-rich stars.  The MR mode has a spectral 
coverage of 0.255\,$\mu$m, and was used to resolve the SiO band heads 
against the continuum in oxygen-rich stars.  The wavelength resolutions ($ 
R= \lambda / \Delta \lambda$) are 360 (1-arcsec slit) and 600 (0.6-arcsec 
slit) in LR mode, and 2000 (1-arcsec slit) and 3300 (0.6-arcsec slit) in MR 
mode.

Telluric standards, which were B dwarfs and giants from the {\it
Hipparcos} catalog, were observed after each target observation.  
The spectrum of the telluric standard is assumed to be a blackbody with 
an effective temperature based on the spectral type from the 
Hipparcos catalog.  Br$\alpha$ and, in some cases, other hydrogen 
recombination lines and the \ion{He}{I} line are detected in the telluric 
standards.
We assume a Gaussian profile for these lines.
Although Starck broadening will affect the profile, 
we ignore this effect because its precise calculation requires a hydrostatic model.
The cancellation of
the atmospheric lines is not good at the  Br$\alpha$
and \ion{He}{I} wavelength in some cases.  The 3.308--3.320\,$\mu$m region of the
spectra is not used in the discussion, because several telluric
methane bands are saturated and cannot be removed.  
 The absolute flux is scaled to $L$-band magnitude,
which is either taken from \cite{Trams99b},
or  estimated from the spectral type, $V$-band magnitude,
and colour of the telluric standard;
it is uncertain to within a factor of two.
The data are
reduced using the {\it eclipse} package and {\it IDL}.  An exposure of
thermal emission from the twilight sky is used for flat-fielding.  
The wavelength calibration is
based on exposures of an Ar+Xe arc lamp with the same wavelength setting 
as the target observations.  The flux error is estimated from ten nearby sky 
pixels in the ISAAC CCD image.

We re-analysed the data for the LMC carbon 
stars whose equivalent
widths were published in \citet{Matsuura02b}.  Only 
the positive
images of chop-and-nod were used in Paper I, while in this 
paper both
the positive and negative images are 
used.

%_________________________________________________________________
\begin{table*}
\begin{caption}
{
Targets and log of the ISAAC observations.
C/O classification shows the basis of classification into oxygen-rich or
carbon-rich (cand: candidate) in publications.
Opt sp: optical spectra, IR sp: infrared spectra (mainly molecule
features), dust: 9.8\,$\mu$m silicate band.
$t_{\mathrm{exp}}$ is the total exposure time in minutes. 
Observing mode shows either LR or MR 
and the slit width in arcsec.
Observing dates are 12th and 14th December, 
2001 for LMC stars, and 24th
July 2002 for the SMC and SgrD.
Ref: reference 
for the 
coordinates.
}
\label{table-target}
\end{caption}
\begin{tabular}{llll  lll l l}\hline\hline 
Name & Alternative name & Coordinates & C/O classification 
& 
$t_{\mathrm{exp}}$ & LR/MR  &  Telluric & ref\\
 & & ($J=2000$) && & + 
Slit & standard
\\
\hline
{\it LMC} \\
{\it Oxygen-rich} \\
\object{IRAS 
04407$-$7000} & LI-LMC 4& 04 40 28.4 $-$69 55 13 &
dust$^{10}$, 
Maser$^{6,14}$, &
60 & MR 0.6 & HIP\,017543 & $^{9,10}$
\\
 & & &  
M7.5$^{16}$
\\
\object{IRAS 04553$-$6825} & WOH SG 64 & 04 55 10.1 $-$68 20 
35 & M7.5:Opt sp$^{1}$, &
33 & MR  0.6 & HIP\,027534 & $^{10,17}$
\\
 & & &  
supergiant$^{1,16}$,
\\
 & & &  Maser$^{6,12,13,14,15,21}$
\\
\object{IRAS 
05042$-$6720} & HV 888   & 05 04 14.3 $-$67 16 17 & Opt
sp M1I$^{16}$, 
M4Ia$^{7}$, &
33 & MR  1.0 & HIP\,027534 & $^{10}$
\\
 & & &  
supergiant$^{8,16}$ 
\\
\object{IRAS 05148$-$6730} & LI-LMC 663 & 05 14 49.9 
$-$67 27 19 & Opt Sp M1I$^{16}$, & 
66 & LR 0.6 & HIP\,025889 & $^{10}$
\\
 
& & &   supergiant$^{8,16}$
\\
\object{IRAS 05558$-$7000} & LI-LMC 1790 & 05 
55 20.8 $-$70 00 05 & dust$^{10}$,Maser$^{6}$&
63 & MR 0.6 & HIP\,029134 & 
$^{9,10}$
\\
% &  HV 916
\\
\multicolumn{4}{l}{\it Oxygen-rich 
(misclassified as carbon-rich before)}\\
\object{IRAS 05128$-$6455} & LI-LMC 
1880 & 05 13 04.6 $-$64 51 40 &
carbon:IR colour, & 
63 & LR 0.6 & 
HIP\,027534 & $^{10}$
\\
 & & &  IR sp$^{10}$
\\
{\it Carbon-rich} 
\\
\object{IRAS 04286$-$6937} & LI-LMC 1825 & 04 28 30.3 $-$69 30 49 & IR 
colour$^{10}$ &
30 & LR 0.6 & HIP\,021949 & $^{9,10}$
\\
\object{IRAS 
04496$-$6958} & LI-LMC 57 & 04 49 18.6 $-$69 53 14 &
Opt sp$^{2}$  &
66 & LR 
  0.6 & HIP\,029635 &$^{9,10}$
\\
& & & Silicate carbon star$^{10}$ 

\\
\object{IRAS 04539$-$6821} & LI-LMC 141 & 04 53 46.3 $-$68 16 12 & IR 
colour$^{10}$ &
63 & LR 0.6 & HIP\,021949 & $^{9,10}$
\\
\object{IRAS 
04557$-$6753} & LI-LMC 198 & 04 55 38.9 $-$67 49 10 & IR colour$^{10}$ &
66 
& LR 0.6 &  HIP\,016368 & $^{9,10}$
\\
\object{IRAS 05112$-$6755} & LI-LMC 
570 & 05 11 10.1 $-$67 52 17 & IR sp/IR colour$^{10}$  &
66 & LR 0.6 & 
HIP\,027534 & $^{9,10}$
\\
%\object{HV 2379}                    & & 05 14 
46.3 $-$67 55 47$^{10}$ & Opt sp? &
\\
\object{SHV 0521050$-$690415}   & & 
05 20 46.8 $-$69 01 25 & Opt sp$^{3}$, IR colour$^{10}$ &
66 & LR 0.6 & 
HIP\,025889 & $^{3,10}$
\\
\object{IRAS 06025-6712}  & LI-LMC 1813 &  
      
                           06 02 31.0 $-$67 12 47 & IR sp$^{11}$ &
36 & LR 
0.6 & HIP\,029727 & $^{9}$
\\
\hline
{\it SMC carbon-rich} \\
\object{   
SMC-S2}                   & & 00 36 59.6 $-$74 19 50 & Opt sp $^{16}$&
48 & 
LR 0.6 & HIP\,012389 & $^{2}$
\\
 & & & Post-AGB cand.$^{2,18}$ \\
\object{  
SMC-S30}                  & & 01 22 29.3 $-$71 09 38 & Opt sp$^{2}$ &
60 & 
LR 0.6 & HIP\,012389 & $^{2}$
 \\
\hline
{\it SgrD carbon-rich}  \\
\object{ 
   Sgr-C1}                   & & 18 46 24.0 $-$30 15 09 & Opt sp$^{5}$   
&
60 & LR 0.6 & HIP\,093542 & $^{19}$
\\
\object{  Sgr-C3}                   
& & 18 52 50.4 $-$29 56 31 & Opt sp$^{5}$   &
32 & LR 0.6 & HIP\,093542 & 
$^{19}$
 \\
\object{  Sgr-UKST3}                & & 18 46 39.1 $-$30 45 52 & 
Opt sp$^{20}$ &
60 & LR 0.6 & HIP\,093542 & $^{19}$
 \\
\object{  
Sgr-UKST15}               & & 18 56 55.9 $-$31 24 40 & Opt sp$^{20}$ &
60 & 
LR 0.6 & HIP\,093542 & $^{19}$
 \\
\hline
\end{tabular}\\
1: \citet{Elias86},
2 : \citet{Groenewegen98},
3 : SHV: \citet{Hughes89},
4 : \citet{Hughes90},
5 : \citet{Ibata95},
6 : \citet{Marshall04},
7 : HV: \citet{Payne-Gaposchkin71},
8 : \citet{Reid90},
9 : LI-LMC: \citet{Schwering90},
10 : \citet{Trams99b},
11 : \citet{vanLoon03},
12 : \citet{vanLoon96},
13 : \citet{vanLoon98a},
14 : \citet{vanLoon98b},
15 : \citet{vanLoon01b},
16 : van Loon (preparation),
17 : WOH: \citet{Westerlund81},
18 : \citet{Whitelock89},
19 : \citet{Whitelock96},
20 : \citet{Whitelock99},
21 : \citet{Wood92}
\end{table*}
%_________________________________________________________________

\section{ Spectra of oxygen-rich stars}

%_________________________________________________________________
\begin{figure}
\centering
%\resizebox{\hsize}{!}{\includegraphics*{o_spec.eps}}
\resizebox{\hsize}{!}{\includegraphics*{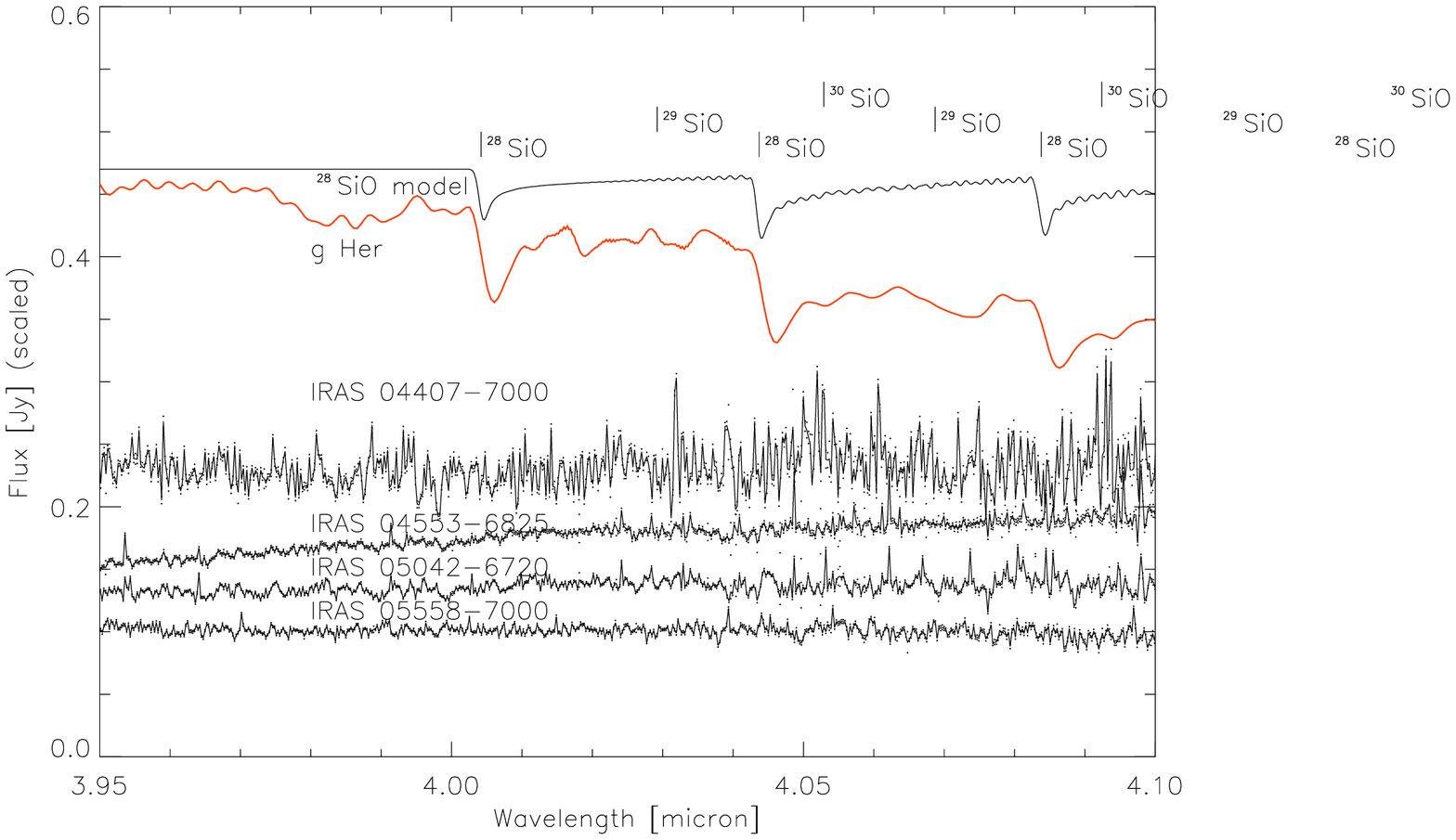}}
\caption{
The 
3.85--4.10\,$\mu$m spectra of four stars LMC oxygen-rich stars.
For 
comparison, a spectrum of the Galactic giant star g\,Her, and a 
model 
spectrum of SiO bands, are also shown.
Wavelength of $^{28}$SiO, $^{29}$SiO, $^{30}$SiO bands are
indicated.
10$\sigma$ errors are plotted as dots.
}
\label{Fig-ospec}
\end{figure}
%_________________________________________________________________

The 3.95--4.10\,$\mu$m spectra of four oxygen-rich stars in the LMC
are shown in Fig.\,\ref{Fig-ospec}, where a model spectrum
of the SiO bands
and a spectrum of a Galactic M giant star (g\,Her)
are plotted as references.  The data of 
g\,Her were taken with ISO/SWS AOT S6
(resolution of about 1600--2000), and the 
spectrum of g\,Her clearly shows
SiO bands.  Parameters for the model 
spectra of SiO  are taken
from \citet{Langhoff93} and \citet{Sauval84}.

The 
oxygen-rich stars in the LMC show no clear features in their spectra.
Although 
the observations targeted the SiO bands, no SiO bands are
detected in any 
of these stars.

The quality of the IRAS 04407$-$7000 spectrum is poor. A 
bump at
4.05\,$\mu$m is caused by insufficient cancellation of 
atmospheric
lines due to Br$\alpha$ absorption in the telluric standard 
star,
and does not imply a Br$\alpha$ emission 
line.

%_________________________________________________________________
\begin{figure}
\centering
%\resizebox{\hsize}{!}{\includegraphics*{IRAS05148_spec.eps}}
\resizebox{\hsize}{!}{\includegraphics*{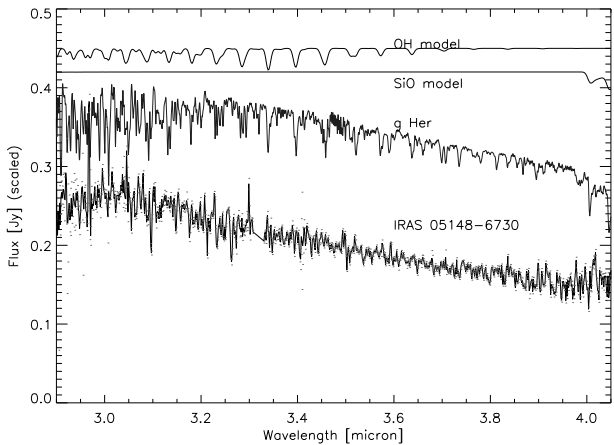}}
\caption{ 
The spectrum of one of the oxygen-rich stars,
IRAS\,05148$-$6730, which 
covers the entire L-band.  For comparison, a
spectra of g\,Her, and OH and 
SiO models are also plotted.  
10$\sigma$ errors are plotted as dots.
}
\label{Fig-05148}
\end{figure}
%_________________________________________________________________

A spectrum covering the full L-band  of the oxygen-rich star
IRAS\,05148-6730 is shown in Fig.\,\ref{Fig-05148}.  A spectrum of
g\,Her and model spectra of OH and SiO lines are also plotted.
Parameters for the model spectra of OH lines are taken from HITRAN
\citep{Rothman98}.  The ISO/SWS spectrum of g Her shows OH, SiO and
possibly weak H$_2$O bands shortwards 3.2\,$\mu$m.
In the spectrum of 
IRAS 05148$-$6730 the decline of the flux shortward of
3.0\,$\mu$m could be due to high-temperature H$_2$O bands;
other than this, no particular molecular bands are detected.  Most of the
features are residual atmospheric lines and noise, rather than OH or SiO
bands. The  H$_2$O continuous
absorption changes its shape depending on the excitation temperature 
and column density. The H$_2$O feature in IRAS 05148$-$6730 will 
not necessarily resemble the one in g\,Her.

\subsection{IRAS\,05128$-$6455}

\begin{figure}
\centering
%\resizebox{\hsize}{!}{\includegraphics*{IRAS05128.eps}}
\resizebox{\hsize}{!}{\includegraphics*{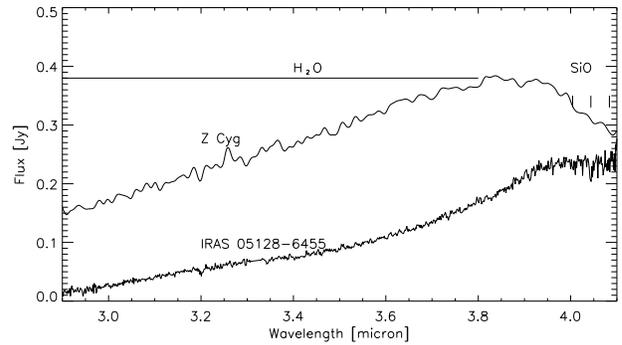}}
\caption{The 
spectrum of IRAS\,05128$-$6455, compared
the spectrum of the 
galactic oxygen-rich Mira, Z\,Cyg
\citep[R$\sim$300;][]{Matsuura02a}, which 
shows H$_2$O and SiO bands.
Deep continuous absorption shortwards of 
3.8\,$\mu$m
in IRAS\,05128$-$6455 is due to H$_2$O. SiO bands may be 
detected,
although the band strength is almost the same as the noise 
level.
A small dip at 3.2\,$\mu$m is the OH 
Q-band.
}
\label{Fig-IRAS05128}
\end{figure}

Fig.\,\ref{Fig-IRAS05128} shows the spectra of 
IRAS\,05128$-$6455 and a
comparison galactic oxygen-rich star 
\object{Z\,Cyg}.  Z\,Cyg shows continuous
H$_2$O absorption up to 
3.8\,$\mu$m, which is also seen in IRAS\,05128$-$6455.
SiO bands are found in 
IRAS\,05128$-$6455 spectra, although the intensity is
almost at noise 
level, due to insufficient cancellation of atmospheric lines. 
 A small dip at 3.20\,$\mu$m  found both in Z\,Cyg and 
in IRAS\,05128$-$6455 is part of the H$_2$O feature with OH blending
 (also found in g\,Her and the OH model spectrum in Fig.\ref{Fig-05148}).

IRAS\,05128$-$6455 was misclassified as a carbon star in
\citet{Trams99b}, 
based on the infrared colours and on the ISO/PHOT
spectrum. The latter 
suggests some absorption-like structure at
3\,$\mu$m, which could be the HCN 
and C$_2$H$_2$ bands.  However, we
could not find such an absorption feature 
in the ISAAC/VLT spectra.
The structure in the ISO/PHOT spectrum appears to 
be the result of an
inaccuracy in the ISO/PHOT detector response curve at 
the time of
reduction.  Such an anomaly is also found in the ISO/PHOT spectrum of
the oxygen-rich star WOH\,G64 \citep{Trams99b}.  
IRAS\,05128$-$6455 is clearly an oxygen-rich star,
as shown by its optical spectrum
(van Loon et al. in preparation).
The spectral type is M9 (van Loon et al.)

%_________________________________________________________________
%_________________________________________________________________

\subsection{Equivalent 
width of the SiO bands}

\begin{figure}
\centering
%\resizebox{\hsize}{!}{\includegraphics*{w40_bw.eps}}
\resizebox{\hsize}{!}{\includegraphics*{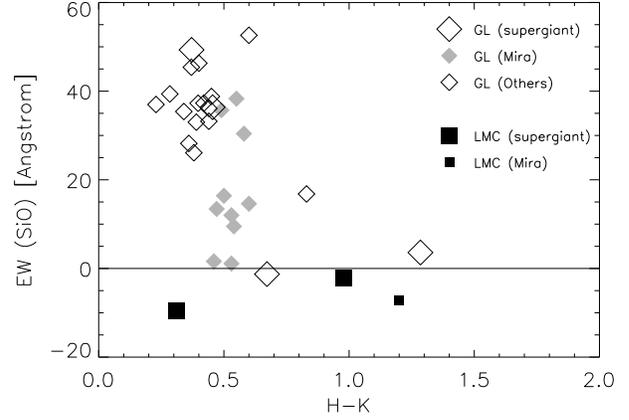}}
\caption{
The 
equivalent width of the SiO bands.
For Galactic stars,
supergiants and Mira 
variables sources are shown by different symbols.
Semi-regular variables 
and unknown variable types are plotted as 
`others'.
}
\label{Fig-eq40}
\end{figure}

The equivalent width (EW) of the SiO bands (\Ws\,) 
are measured for
LMC oxygen-rich stars and compared with the galactic 
sample
(Fig.\ref{Fig-eq40}, Table\,\ref{table-oxygen}).  The definition 
of
\Ws\, is similar to the one in \citet{Aringer97, Aringer99}, who
studied 
SiO bands in Galactic AGB stars.  The \Ws\, is the sum of
equivalent widths 
of three strong $^{28}$SiO bands.  The wavelengths
used for the continuum and 
the SiO bands are the same as
\citet{Aringer97, Aringer99}.  In 
\citet{Aringer97}, the continuum is
estimated by a polynomial fit.  Here we use a linear fit, because
third-order polynomials were 
inappropriate and  because
we occasionally have insufficient data to attempt higher order fits.
This 
difference of fitting has little influence on the measured EW, as
tested on 
SWS spectra: less than 5\,\AA. The exception is KK\,Per, for
which we found a difference of about 10\,\AA.  The precise wavelength
regions used for the 
continuum are summarized in Table\,\ref{table-def}.

The one-$\sigma$ error 
in the EW is propagated from the noise in the
spectrum, and is less than 
1\,\AA.  However, in practice, the error of
the local continuum estimate has 
a larger uncertainty, as mentioned
above. \citet{Aringer99} found a 
similar uncertainty up to
5\,\AA.

To allow comparison with the LMC 
sample, we used the SiO EWs of
Galactic AGB stars given in 
\citet{Aringer99}.  In addition, we reduced
ISO/SWS spectra that were 
observed in AOT 6 (wavelength resolution of
about {\it R}=1600--2000), and AOT 1 speed 4
({\it R}=1000--1400) which was used for $\alpha$~Ori.
We also use reduced ISO/SWS AOT 6 spectra from
\citet{Sloan03}.

The measured EW of SiO 
bands are summarized in
Table\,\ref{table-oxygen}.  No evidence for 
absorption bands is found,
consistent with a non-detection of SiO.  Three 
LMC stars are affected
by data problems and are listed separately; for 
IRAS\,05128$-$6455 and
IRAS\,05148$-$6730 the wavelength resolution is not 
high enough to
separate the continuum and the absorption, while IRAS 
04407$-$7000 has
a poor quality spectrum.

Fig.\ref{Fig-eq40} shows the 
\Ws\, of LMC and Galactic oxygen-rich
stars, as a function of $H-K$ colour.  
Here we plot the \Ws\, of three
LMC stars only, avoiding any with poor quality data
The $H-K$ colour is the
measure of the effective temperature for blue 
stars, but not for red
stars where the colour is strongly affected by dust.  For both LMC
and Galactic stars, large \Ws\ stars (larger than 20 
\AA) are absent
among red stars ($H-K>0.7$).  Within our sample of blue 
Galactic
stars, there is no difference in \Ws\ between supergiants and 
giants
(Mira and semi-regular variables) as is consistent with 
previous
observations \citep{Wallace02, 
Heras02}.

%_________________________________________________________________
\begin{table}
\begin{caption}
{
The 
definition of equivalent width ($EW$) for the molecular absorption 
bands.
The continuum level is linearly interpolated from the 
wavelength
region in the last column.
SiO molecules show three strong bands 
from 4.0--4.1\,$\mu$m,
and \Ws\, is the total of these three 
bands.
}\label{table-def}
\end{caption}
\begin{tabular}{llll}
\hline\hline
Name 
& Molecules  & Measured        & Continuum              \\
     &            
& region [$\mu$m] & [$\mu$m]               \\
\hline
\Wo  & HCN, \CtHt & 
2.97--3.35      & 2.94--2.97, 3.35--3.40 \\
\Wf  & HCN        & 3.56--3.58   
    & 3.51--3.56, 3.58--3.63 \\
\We  & \CtHt      & 3.60--4.00      & 
3.50--3.60, 4.00--4.10 \\
\hline
\Ws  & SiO        & 4.0025--4.0200  & 
3.95--3.96, 3.99--4.00 \\
     &            & 4.0425--4.0600  & 4.036--4.04  
           \\
     &            & 4.0825--4.1000  & 4.076--4.080           
\\
\hline
\end{tabular}
\end{table}
%_________________________________________________________________
%_________________________________________________________________
\begin{table*}
\begin{caption}
{The 
equivalent width of SiO bands (unit is \AA). 
The date is that of either ISAAC or ISO/SWS observations in $ddmmyy$
format, which is different from the observing date of the magnitudes.
Column four
($\phi _{\rm{K}}$) lists
the pulsation phase during the SiO observations,
which is estimated from infrared K-band light curve
\citep{Whitelock03}, where 0.0 and 1.0
are the maxima. Uncertainty 
should be about 0.1.
The optical phase ($\phi_{\mathrm opt}$) is estimated for the ISO sample,
based on the AAVSO light curve.
The optical phase is usually about 0.1--0.2 of a cycle ahead of the infrared phase.
Photometric magnitudes are taken from \citet{Trams99b} 
for the LMC sample,
and 2MASS for the Galactic sample.
Luminosity class and 
spectral types  for Galactic
stars are taken from 
Simbad.
}
\label{table-oxygen}
\end{caption}
\begin{tabular}{ll cll r@{.}l | 
r@{.}l r@{.}l r@{.}l r@{.}l r@{.}l r@{.}l r@{.}l llll}\hline\hline 
Name              &  Sp Type & date & $\phi _{\rm{K}}$ &
$\phi_{\mathrm opt}$ & \multicolumn{2}{c}{\Ws} & 
 \multicolumn{2}{c}{H}  &    \multicolumn{2}{c}{K}  
\\
\hline
%\it LMC sample
\multicolumn{6}{l}{{\it LMC sample}} &
\\
IRAS 04553$-$6825 & M7.5, supergiant      &121201&      && $-$2&2  &    7&96 &   6&98 \\
IRAS 05042$-$6720 & M1I, M4Ia, supergiant &121201&      && $-$9&7  &    7&20 &   6&89 \\
IRAS 05558$-$7000 &                       &131201& 0.87 && $-$7&3  &   10&10 &   8&90 \\
\hline
\multicolumn{6}{l}{{\it LMC sample (Poorer quality in \Ws)}}&\\
IRAS 04407$-$7000 &                       &141201& 0.82 &&$-$24&2  &    8&92 &   8&18 \\
IRAS 05128$-$6455 &                       &131201&      &&   6&0   &   12&10 &  10&55 \\
IRAS 05148$-$6730 & M1I, supergiant       &151201&      &&   17&5 \\
\hline
\multicolumn{6}{l}{{\it Galactic sample from ISO}}&\\
R Aql        & Mira                &190397&&0.25  &   13&4       &$-$0&355$\pm$0.218 & $-$0&826$\pm$0.230 \\
VY CMa       & M5Ia,e              &211197&&      &    3&6       &   1&576$\pm$0.210 &    0&291$\pm$0.236 \\
g Her        & M6III, SR           &230896&&      &   39&4       &$-$1&850$\pm$0.138 & $-$2&134$\pm$0.158 \\
$\alpha$ Ori & M1Iab, SR           &081097&&      &   49&3       &$-$4&007$\pm$0.162 & $-$4&378$\pm$0.186 \\
$\rho$ Per   & M4II, SR            &050398&&      &   37&0       &$-$1&675$\pm$0.158 & $-$1&904$\pm$0.152 \\
KK Per       & M2Iab-Ib            &030298&&      &   36&4       &   2&138$\pm$0.166 &    1&683$\pm$0.206 \\
VX Sgr       & M5/M6III, M4eIa, SR &131096&&      & $-$1&3       &   0&550$\pm$0.304 & $-$0&122$\pm$0.362 \\
SW Vir       & M7III, SR           &200796&&      &   37&3       &$-$1&606$\pm$0.260 & $-$2&003$\pm$0.336 \\
\hline
\end{tabular}\\
\end{table*}
%______________________________________________________________
%_________________________________________________________________

\section{ Spectra of carbon-rich stars}\label{sect-cstar}

%_________________________________________________________________
\begin{figure*}
\centering
%\resizebox{\hsize}{!}{\includegraphics*{c_spec.eps}}
\resizebox{\hsize}{!}{\includegraphics*{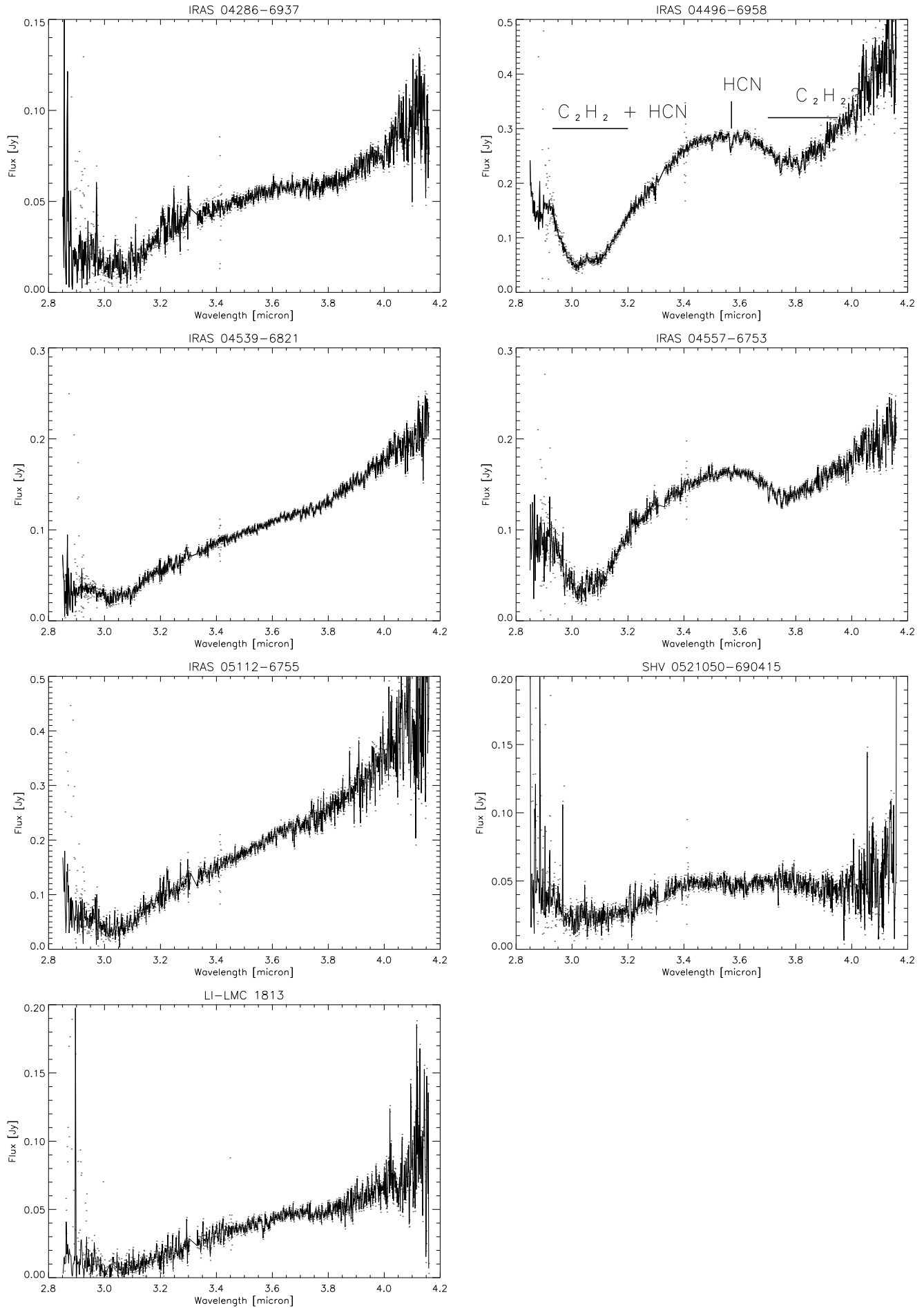}}
\caption{
ISAAC 
spectra of carbon-rich AGB stars in LMC.
Molecule bands, in particular the 
3.1\,$\mu$m C$_2$H$_2$+HCN,
3.5\,$\mu$m HCN, and 3.8 \,$\mu$m, probably 
C$_2$H$_2$ bands,
are indicated.
10$\sigma$ errors are plotted as dots.
}
\label{Fig-cspec}
\end{figure*}
%_________________________________________________________________
%_________________________________________________________________
\begin{figure*}
\centering
%\resizebox{\hsize}{!}{\includegraphics*{c_smc_spec.eps}}
\resizebox{\hsize}{!}{\includegraphics*{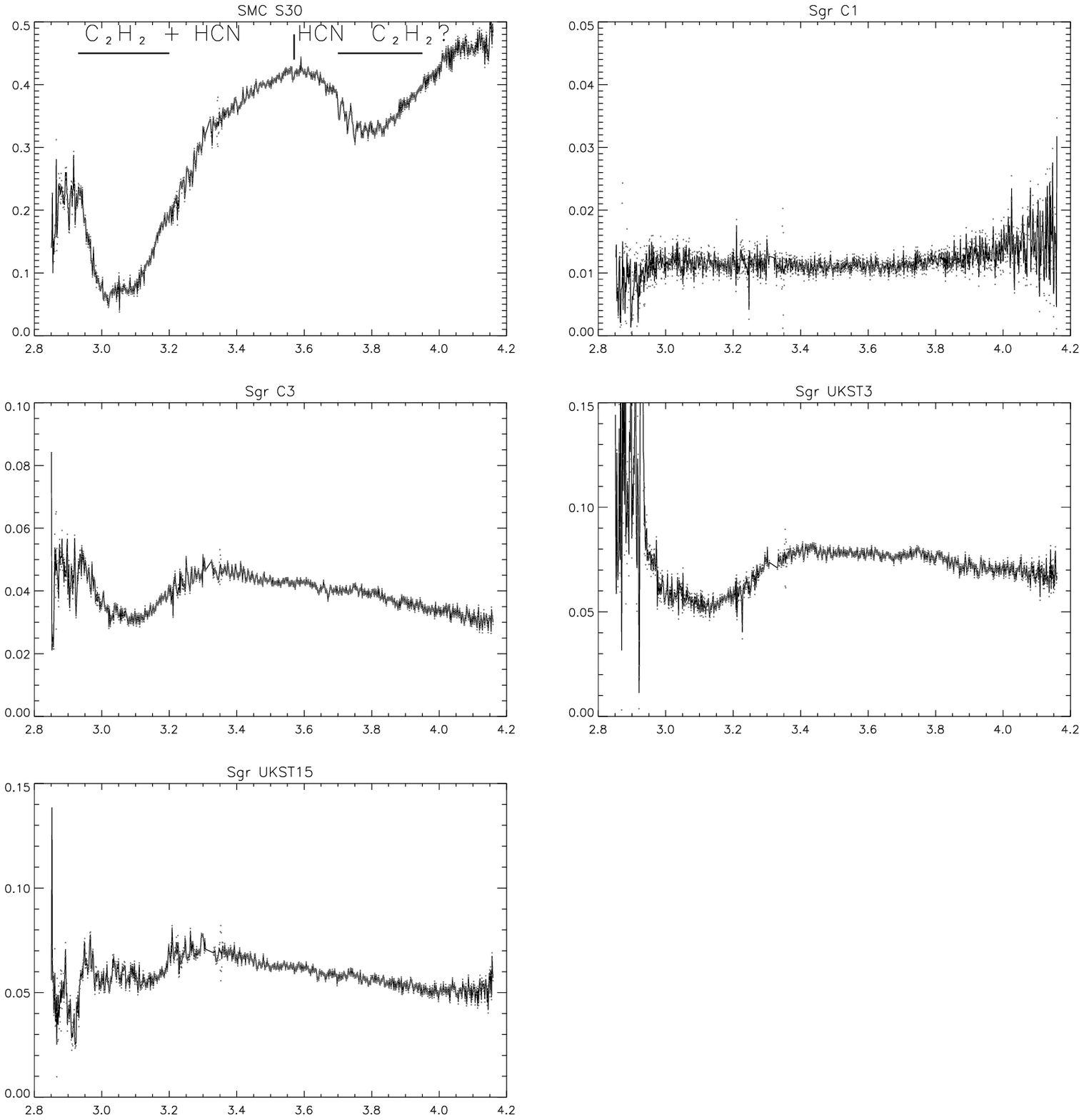}}
\caption{
Same 
as Fig.\,\ref{Fig-cspec}, but for carbon-rich stars in SMC and 
SgrD.
}
\label{Fig-cspec-smc}
\end{figure*}
%_________________________________________________________________

Fig.\,\ref{Fig-cspec} 
shows the ISAAC spectra of carbon-rich stars in the LMC, while spectra of 
carbon-rich AGB stars in the SMC and SgrD are shown in 
Fig.\,\ref{Fig-cspec-smc}.  Bands of different molecules are found in the 
L-band: the 3.1\,$\mu$m HCN+C$_2$H$_2$ band, 3.4 and 3.7\,$\mu$m CH, 
3.5\,$\mu$m HCN, and 3.8\,$\mu$m C$_2$H$_2$. The last identification is 
probable rather than certain.

The 3.1\,$\mu$m HCN+C$_2$H$_2$ is commonly seen in carbon-rich stars,
but is more pronounced in bluer stars, such as IRAS\,04496$-$6958.
The sharp absorption at 3.57\,$\mu$m is due to HCN Q-branch lines.
The obvious features are blend of lines from (00$^0$0) to (01$^1$1)
and some other transitions \citep{Harris03}.  Lines with other
transitions of the HCN Q-branch, around 3.5\,$\mu$m, are not obvious in
our spectra.  The 3.5\,$\mu$m feature is found in only a few of the
extra-galactic stars: IRAS 04496$-$6958, IRAS 06025$-$6712, SMC S30, and 
probably IRAS 04557$-$6753 and IRAS 05112$-$6755.

A non-detection of the 3.5\,$\mu$m HCN band does not immediately mean
no contribution by HCN to the 3.1\,$\mu$m band.  The absorbance ratio
of the 3.1 $\mu$m /3.5 $\mu$m HCN bands is approximately 100
\citep*{Harris02}, and the 3.1\,$\mu$m feature is easily formed even
at lower HCN column density.

The identification of the C$_2$H$_2$ feature is still tentative.
\citet{Goebel81} noted the 3.8\,$\mu$m feature on a KAO spectrum of
the carbon star \object{V CrB}.  \citet{Hron98}'s model calculation
shows a C$_2$H$_2$ band at 3.8\,$\mu$m, and they identify this with a
feature in the ISO/SWS spectra of the Galactic carbon star R\,Scl.
However, the absorption in the R\,Scl spectrum starts 
only at 3.9\,$\mu$m, which could indicate CS \citep{Aoki98}. The
identification of the 3.8 \,$\mu$m band, which stretches from 3.6 to
4.0\,$\mu$m, as a C$_2$H$_2$ feature is still not fully fixed.

\begin{figure}
\centering
%\resizebox{\hsize}{!}{\includegraphics*{ch.ps}}
\resizebox{\hsize}{!}{\includegraphics*{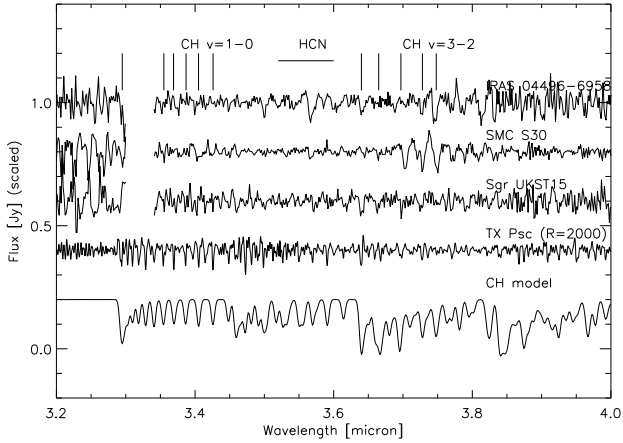}}
\caption{ 
The flattened spectra of the carbon-rich stars, compared
with the CH model 
spectrum and with the ISO/SWS spectrum of TX Psc.
The CH line list is taken 
from \citet{Jorgensen96}.  The relative
intensities of the CH lines in the 
model are not always comparable to
the observed ones, especially around 3.5 
and 3.8 \,$\mu$m, because of
other opacity sources such as the HCN and 
C$_2$H$_2$ bands.  
}
\label{Fig-ch}
\end{figure}

There are several other weak lines visible in the 
spectra around
3.4\,$\mu$m and at 3.7\,$\mu$m, where  weak lines are seen 
superposed on the broad C$_2$H$_2$ band.  These lines are due to CH
(Fig.\,\ref{Fig-ch}),  whose bands are found in the Galactic carbon star
\object{TX Psc} \citep{Ridgway84, Aoki98}: its ISO/SWS spectrum
\citep{Aoki98} is added in Fig.\,\ref{Fig-ch} for comparison.
The resolution of the TX Psc spectrum is $R\sim2000$, higher than
the ISAAC LR spectra ($R=600$), and the CH bands appear more clearly.  In the
ISAAC spectra, the CH $v$=1--0, and $v=$3--2 bands are easy to find
while $v=$2--1 at 3.5 \,$\mu$m is not obvious.  This could be
because of the blend with the 3.5\,$\mu$m HCN bands.  Note that
C$_2$H$_2$, which contains C--H bonds, may also contribute narrow
features at 3.8$\mu$m, but these are not definitely detected.

Sgr-C1 shows a flat spectrum without any feature in the L-band: no bands 
are visible.  The flat spectrum (in Jy) at L implies a red colour which shows 
that this  is a late-type star, but we were unable to confirm that it is actually a
carbon star.  Sgr-C1 is one of the first stars found in SgrD \citep{Ibata95},
and it is an optically classified carbon star.  It is unlikely that we took
the spectrum of the wrong object, because 
the VLT pointing is in general accurate enough and the acquisition
image of ISAAC is consistent with the finding chart in
\citet{Whitelock96}.  \citet{Whitelock99} show 
that the K-band magnitude of Sgr-C1 is 11.2\,mag, fainter than the  
three other Sgr-D carbon-rich stars (9.4, 9.8 and 9.7\,mag).  The luminosity of 
Sgr-C1 is $M_{\rm K}\sim -6.1$\,mag, which is not near the tip of the AGB
sequence \citep{Whitelock96}. All these suggest that Sgr-C1 could be
an extrinsic carbon star.

%_________________________________________________________________
%_________________________________________________________________

\subsection{ The post-AGB star, SMC-S2}

\begin{figure}
\centering
%\resizebox{\hsize}{!}{\includegraphics*{smc_s2.eps}}
\resizebox{\hsize}{!}{\includegraphics*{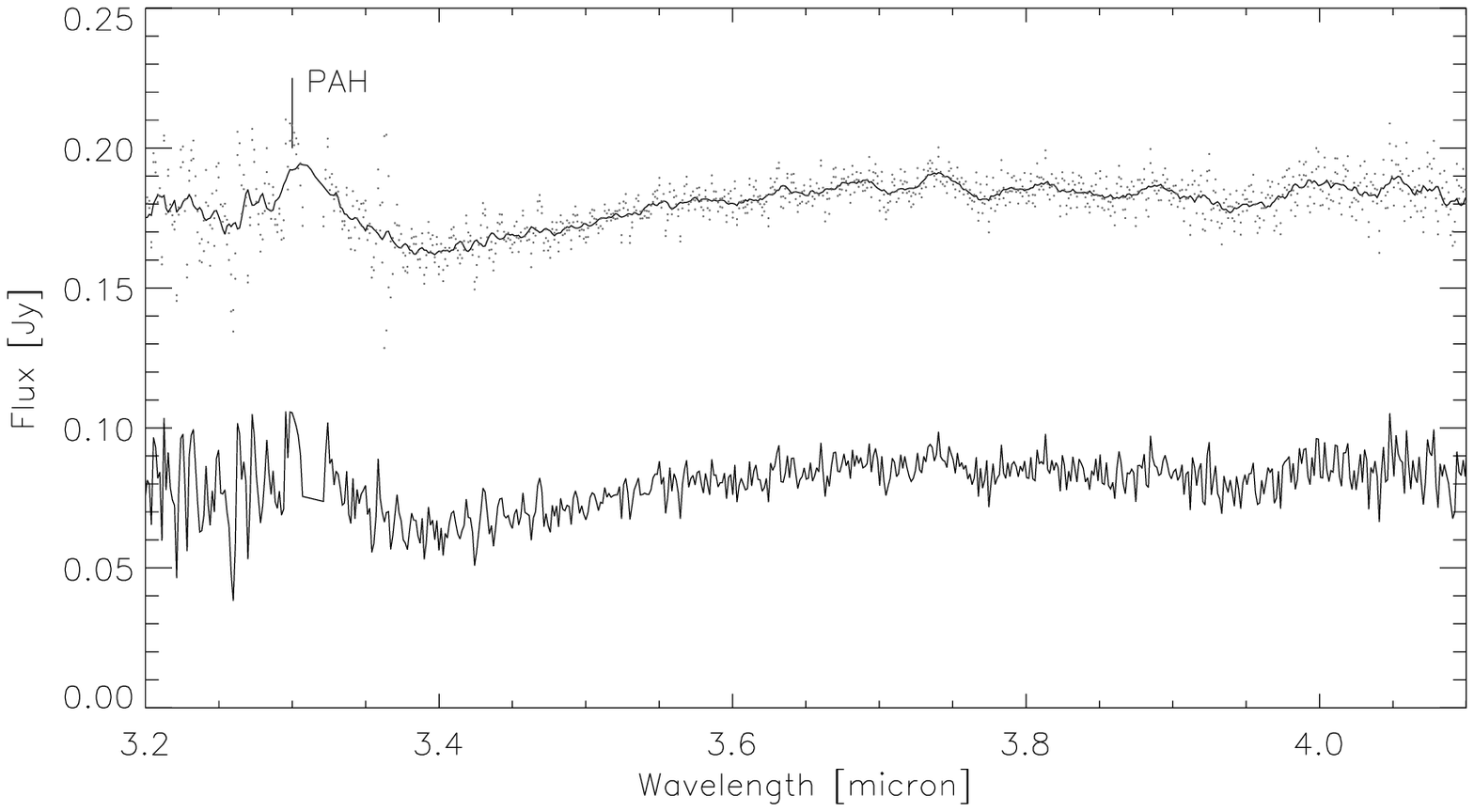}}
\caption{
The spectrum of SMC-S2, and smoothed spectrum to 1/10th of the
original wavelength resolution.
The 3.3\,$\mu$m PAH band is detected. (The 
3.308--3.320\,$\mu$m region is removed because of strong telluric lines.)
10$\sigma$ errors are plotted as dots.
}
\label{Fig-cspec-postagb}
\end{figure}

In the spectrum of SMC-S2, the 3.3\,$\mu$m PAH 
band is detected (Fig.\,\ref{Fig-cspec-postagb}). Although this confirms 
its carbon-rich
nature, it also shows that this object is unlikely to be an AGB star,
as PAH emission is never seen in AGB stars. The temperatures of AGB
stars are too low to excite PAHs: this requires UV or optical photons.
PAH emission is usually seen once the stars reach a temperature of $\sim
5000\,$K. The presence of PAH emission indicates that SMC-S2 is a
post-AGB star, which was already suspected from 
its infrared colours and the optical spectra \citep{Whitelock89,
Groenewegen98}. This is probably the first detection of a PAH band in
an SMC post-AGB star.

The optical spectrum suggests a K-type star. This is rather cool for
PAH excitation but the detected feature is weak. It is possible that the 
underlying star is hotter  than implied by the spectral type
and that the spectrum is masked by the stellar 
envelope. \citet{Whitelock89} discuss whether the object may be an 
interacting binary. This cannot be excluded, but the weak variability 
and high luminosity are consistent with a post-AGB nature, and there is no 
evidence for ionization in the spectrum.

PAH emission is widespread in the 
interstellar medium (ISM) of the
SMC, as shown by ISOCAM observations 
\citep{Bot04}, but at a low
abundance relative to the gas as expected from 
its metallicity.  The
extinction curve in the bar of the SMC rises very 
steeply to the UV
\citep{Gordon03}, which can be attributed to PAHs.  It is not
known whether the PAHs form in the ISM or derive from the same 
sources as
the dust.  PAH detection in SMC-S2 shows that PAHs can
form in the metal-poor stars, leaving evolved stars as a potential
source.

%_________________________________________________________________
%_________________________________________________________________

\subsection{ 
Equivalent widths of the molecular 
bands}

%_________________________________________________________________
\begin{table*}
\begin{caption}
{Equivalent widths of molecules found in carbon-rich stars.
 Date ({\it ddmmyy}) shows the observing date of ISAAC or ISO/SWS, which is different
from that of photometric observations.
}
\label{table-carbon}
\end{caption}
\begin{tabular}{lll  r@{.}l  r@{.}l r@{.}l | r@{.}l r@{.}l r@{.}l }\hline\hline 
Name         &  date & $\phi_{opt}$ &
\multicolumn{2}{c}{$\Wo$}        &
\multicolumn{2}{c}{$\Wf$}        & 
\multicolumn{2}{c}{$\We$}        & 
\multicolumn{2}{c}{H }&  
\multicolumn{2}{c}{K}   \\ 
 & & &
\multicolumn{2}{c}{($\times10^{-2}$)} & \multicolumn{2}{c}{} & 
\multicolumn{2}{c}{($\times10^{-2}$)} \\
 & & & \multicolumn{2}{c}{[\AA]} & 
\multicolumn{2}{c}{[\AA]} &
\multicolumn{2}{c}{[\AA]} & 
\multicolumn{2}{c}{[mag]} &
\multicolumn{2}{c}{[mag]}
\\
\hline
\multicolumn{8}{l}{\it LMC} &  \\
  IRAS 04286$-$6937    &141201&&  7&00$\pm$ 0.08 &$-$2&10$\pm$ 0.87 &  3&626$\pm$ 0.003 & 13&00$\pm$0.05 &  11&25$\pm$0.05   \\
  IRAS 04496$-$6958    &121201&& 12&26$\pm$ 0.02 &   7&82$\pm$ 0.63 &  7&167$\pm$ 0.005 & 10&90$\pm$0.05 &   9&50$\pm$0.04   \\
  IRAS 04539$-$6821    &131201&&  6&55$\pm$ 0.03 &   1&37$\pm$ 0.48 &  1&803$\pm$ 0.002 & 14&30$\pm$0.05 &  11&80$\pm$0.04   \\
  IRAS 04557$-$6753    &131201&& 10&84$\pm$ 0.03 &   1&08$\pm$ 0.42 &  5&556$\pm$ 0.002 & 14&60$\pm$0.05 &  12&40$\pm$0.04   \\
  IRAS 05112$-$6755    &151201&&  8&37$\pm$ 0.04 &$-$0&01$\pm$ 0.69 &  2&500$\pm$ 0.004 & 14&70$\pm$0.10 &  12&00$\pm$0.05   \\
  SHV 0521050$-$690415 &151201&& 13&12$\pm$ 0.06 &  12&87$\pm$ 1.22 &  1&068$\pm$ 0.009 &  9&77$\pm$0.02 &   9&23$\pm$0.02   \\
  LI$-$LMC 1813        &131201&& 11&10$\pm$ 0.18 &  14&05$\pm$ 0.22 &  2&065$\pm$ 0.005  \\
\hline
\multicolumn{8}{l}{\it SMC} & \\  
  SMC-S30              &240702&& 14&58$\pm$ 0.00 &$-$0&79$\pm$ 0.16 &  5&944$\pm$ 0.001 &  13&48$\pm$0.03 &   11&42$\pm$0.03$^4$ \\     
\hline
\multicolumn{8}{l}{\it SgrD} & \\
  Sgr-C3               &240702&&  6&96$\pm$ 0.01 &  1&29$\pm$  0.28 &  0&033$\pm$ 0.002 &  10&02 &    9&44$^5$ \\
  Sgr-UKST3            &240702&&  7&94$\pm$ 0.02 &  1&83$\pm$  0.36 &  0&017$\pm$ 0.002 &  10&13 &    9&51$^5$ \\
  Sgr-UKST15           &240702&&  2&54$\pm$ 0.01 &  0&74$\pm$  0.38 &  1&360$\pm$ 0.002 &  10&08 &    9&65$^5$ \\
\hline
\multicolumn{8}{l}{\it Galactic carbon-rich stars} & \\
\object{LP And}         &021296 &    &   3&44 &   3&85 & $-$0&46 &   6&355$\pm$0.018$^7$ &  3&859$\pm$0.036$^{7,*}$  \\
\object{VX And}         &170197 &    &  12&42 &$-$0&21 &    0&07 &   1&891$\pm$0.202     &  1&193$\pm$0.202$^7$\\  
\object{V Aql}          &290496 &    &   8&71 &   1&10 &    0&48 &   0&64                &$-$0&09              \\
\object{WZ Cas}         &220796 &    &  10&84 &   9&85 & $-$0&37 &   1&01                &   0&57$^3$          \\
\object{S Cep}          &310597 &    &   6&44 &   3&60 &    2&07 &   1&013$\pm$0.186     &$-$0&029$\pm$0.192$^7$\\  
\object{V CrB}          &070396 &0.20&   8&75 &  10&37 &    3&08 &   2&209$\pm$0.218     &   1&321$\pm$0.276$^7$\\  
                        &290796 &0.17&   8&56 &   9&90 &    3&07 &   \multicolumn{2}{c}{}&   \multicolumn{2}{c}{}\\  
                        &110197 &0.64&   7&77 &   6&28 &    2&00 &   \multicolumn{2}{c}{}&   \multicolumn{2}{c}{}\\  
                        &060397 &0.19&   5&16 &   2&01 &    0&35 &   \multicolumn{2}{c}{}&   \multicolumn{2}{c}{}\\  
                        &120697 &0.78&   7&32 &   4&12 &    1&45 &   \multicolumn{2}{c}{}&   \multicolumn{2}{c}{}\\  
                        &210997 &0.32&   8&88 &   9&22 &    2&18 &   \multicolumn{2}{c}{}&   \multicolumn{2}{c}{}\\  
\object{Y CVn}          &250496 &    &   8&77 &\multicolumn{2}{c}{}&1&13 &$-$0&285$\pm$0.166 &$-$0&738$\pm$0.166$^7$\\  
\object{V Cyg}          &050296 &0.52&   9&26 &  10&05 &    3&44 &   1&273$\pm$0.198     &   0&117$\pm$0.192$^7$ &  \\
\object{V460 Cyg}       &110197 &    &   4&66 &\multicolumn{2}{c}{}&1&14 &   0&583$\pm$0.192 & 0&270$\pm$0.184$^7$\\  
                        &291197 &    &   3&02 &\multicolumn{2}{c}{}&1&00 & \multicolumn{2}{c}{}   & \multicolumn{2}{c}{}\\  
\object{RY Dra}         &120597 &    &   7&74 &   1&38 &    0&72 &   0&955$\pm$0.214     &   0&394$\pm$0.204$^7$\\  
\object{T Dra}          &070396 &0.59&   7&29 &  12&86 &    4&33 &   2&454$\pm$0.192     &   1&365$\pm$0.212$^7$\\  
                        &210796 &0.93&   7&81 &   8&21 &    3&09 &   \multicolumn{2}{c}{}&   \multicolumn{2}{c}{}\\  
                        &281096 &0.17&   6&98 &   6&78 &    2&27 &   \multicolumn{2}{c}{}&   \multicolumn{2}{c}{}\\  
                        &041296 &0.25&   7&51 &  11&69 &    3&74 &   \multicolumn{2}{c}{}&   \multicolumn{2}{c}{}\\  
                        &180197 &0.36&   5&67 &   4&68 &    1&64 &   \multicolumn{2}{c}{}&   \multicolumn{2}{c}{}\\  
                        &260197 &0.38&   6&15 &   4&46 &    1&81 &   \multicolumn{2}{c}{}&   \multicolumn{2}{c}{}\\  
                        &150597 &0.64&   7&77 &   7&88 &    2&89 &   \multicolumn{2}{c}{}&   \multicolumn{2}{c}{}\\  
                        &210897 &0.86&   5&65 &   8&87 &    2&80 &   \multicolumn{2}{c}{}&   \multicolumn{2}{c}{}\\  
\object{R For}          &130298 &    &   6&83 &   0&91 &    1&85 &   2&537$\pm$0.206     &   1&349$\pm$0.274$^7$\\  
\object{CW Leo}         &310596 &    &   1&81 &   1&05 & $-$0&33 &   4&25                &   1&28$^2$           \\
\object{W Ori}          &220398 &    &   4&90 &   0&01 &    0&26 &   0&092$\pm$0.338     &$-$0&470$\pm$0.402$^7$\\  
\object{RZ Peg}         &100696 &    &   9&08 &  13&15 &    1&98 &   2&850$\pm$0.214     &   2&127$\pm$0.302$^7$\\  
\object{TX Psc}         &261196 &    &   5&34 &   2&35 &    0&87 &$-$0&33                &$-$0&67$^3$           \\
\object{R Scl}          &281196 &0.28&  13&47 &   6&92 &    2&09 &   0&75                &$-$0&07$^2$           \\
\object{S Sct}          &290496 &    &   9&06 &   4&18 &    2&06 &   1&06                &   0&44$^3$           \\
\object{SZ Sgr}         &280398 &    &   3&72 &   \multicolumn{2}{c}{}     &    1&19     &   2&623$\pm$0.216 & 2&168$\pm$0.234$^7$\\ 
\object{TU Tau}         &180398 &    &   4&78 &   1&67 &    1&16 &   2&093$\pm$0.210     &   1&574$\pm$0.220$^7$\\  
\object{RU Vir}         &200796 &    &   4&86 &   6&88 &    1&50 &   2&50                &   1&33$^3$           \\
\object{SS Vir}         &140696 &    &  14&80 &   6&80 &    3&80 &   1&811$\pm$0.234     &   0&831$\pm$0.330$^7$\\  
\object{AFGL 940}       &040498 &    &   1&75 &   2&32 & $-$0&16 &   6&269$\pm$0.013     &   4&391$\pm$0.011$^7$\\  
\object{AFGL 2155}      &010397 &    &   3&38 &   3&05 &    0&51 &   8&722$\pm$0.038     &   5&609$\pm$0.018$^7$\\
\object{AFGL 2392}      &220398 &    &   4&87 &   7&32 &    0&34 &   5&24                &   3&65$^2$           \\
\object{CS 2178}        &250197 &    &   7&52 &   \multicolumn{2}{c}{}     &    2&05     &   4&158$\pm$0.172 & 2&798$\pm$0.234$^7$\\   
\object{CS 2429}        &200297 &    &   6&18 &   \multicolumn{2}{c}{}     &    0&19     &   1&699$\pm$0.342 & 1&007$\pm$0.346$^7$\\  
\object{DO 40123}       &150197 &    &  11&74 &   8&69 &    3&35 &   1&867$\pm$0.182     &   0&913$\pm$0.192$^7$\\  
\object{IRC$-$10095}    &010498 &    &  13&94 &   1&23 &    2&84 &   3&235$\pm$0.206     &   2&241$\pm$0.248$^7$\\  
\object{IRC$-$10122}    &310398 &    &   6&01 &   2&07 &    1&25 &   3&851$\pm$0.248     &   2&399$\pm$0.272$^7$\\  
\object{IRC+00365}      &290397 &    &   7&17 &   4&49 &    1&69 &   4&100$\pm$0.182     &   2&209$\pm$0.242$^7$\\  
\object{IRC+50096}      &020298 &    &   5&22 &   5&15 &    1&69 &   2&596$\pm$0.192     &   1&150$\pm$0.192$^7$\\  
\object{IRAS19068+0544} &090397 &    &   6&06 &   \multicolumn{2}{c}{}     &    2&51     &   4&685$\pm$0.075 & 3&080$\pm$0.254$^7$\\ 
\hline
\end{tabular}\\ 
1: \citet{Gezari93},
2: \citet{LeBertre92} (average over the all phase),
3: \citet{Noguchi81},
4: \citet{Groenewegen98},
5: \citet{Whitelock96}
6: \citet{Trams99b},
7: 2MASS,
$^*$: 2MASS quality flag is `E', i.e., poor quality
\end{table*}
%_________________________________________________________________

%_________________________________________________________________
\begin{figure}
\centering
%\resizebox{\hsize}{!}{\includegraphics*{w35_bw.eps}}
\resizebox{\hsize}{!}{\includegraphics*{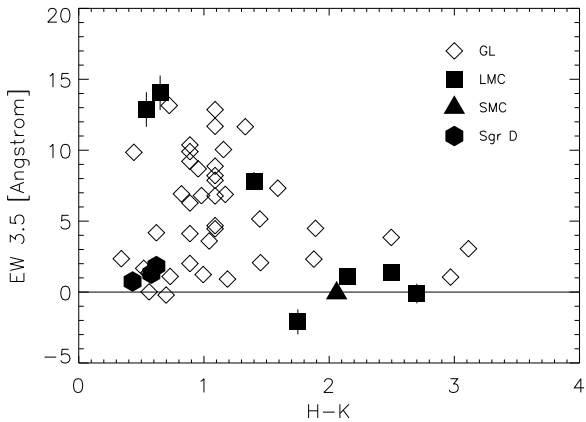}}
\caption{
The 
HCN equivalent widths of \Wf.
The symbols show the host galaxies of the 
samples.
}
\label{Fig-eq35}
\end{figure}
%_________________________________________________________________
%_________________________________________________________________
\begin{figure}
\centering
%\resizebox{\hsize}{!}{\includegraphics*{w38_bw.eps}}
\resizebox{\hsize}{!}{\includegraphics*{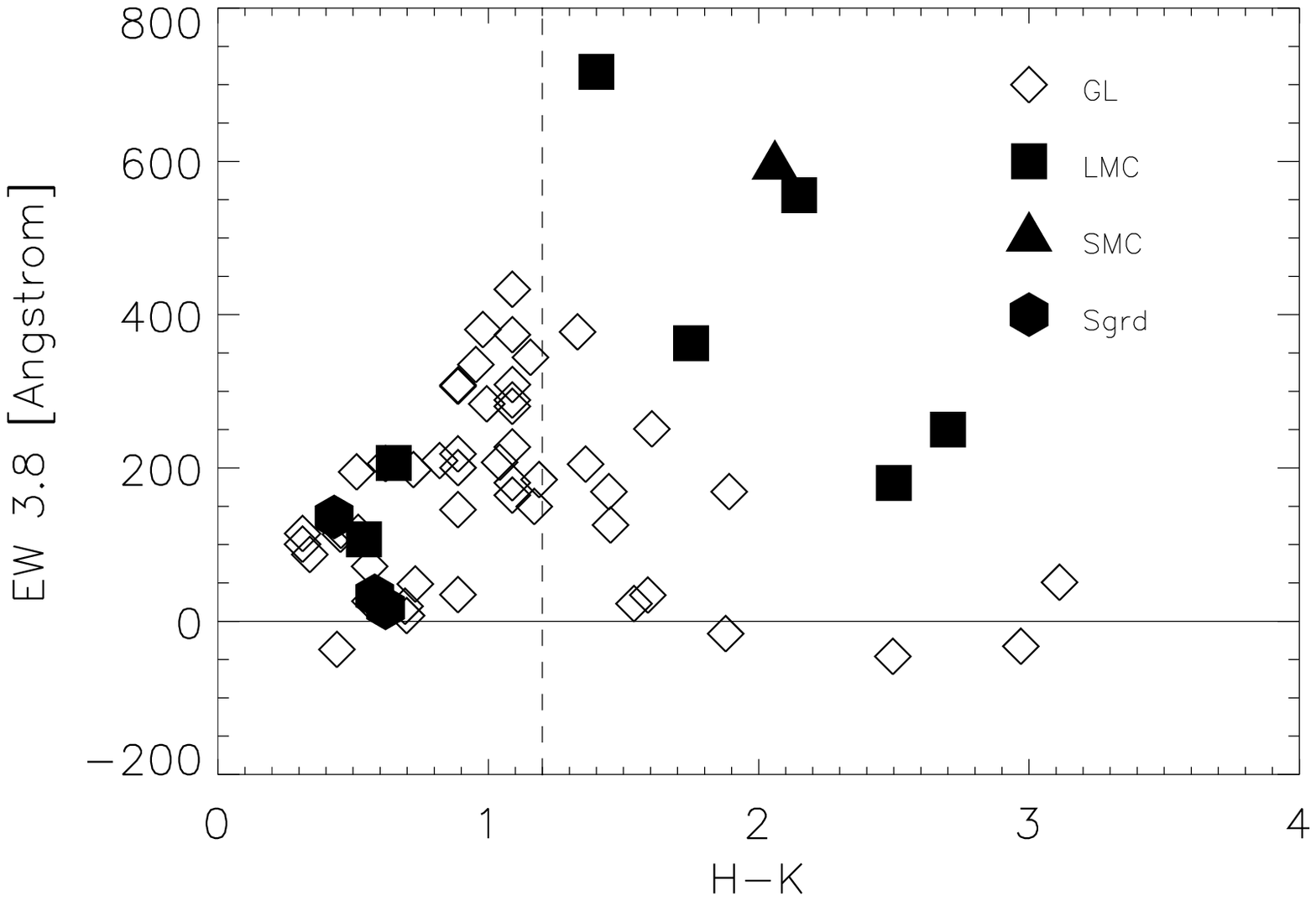}}
\caption{
The 
C$_2$H$_2$ equivalent widths of 
\We.
}
\label{Fig-eq38}
\end{figure}
%_________________________________________________________________
%_________________________________________________________________
\begin{figure}
\centering
%\resizebox{\hsize}{!}{\includegraphics*{w31_bw.eps}}
\resizebox{\hsize}{!}{\includegraphics*{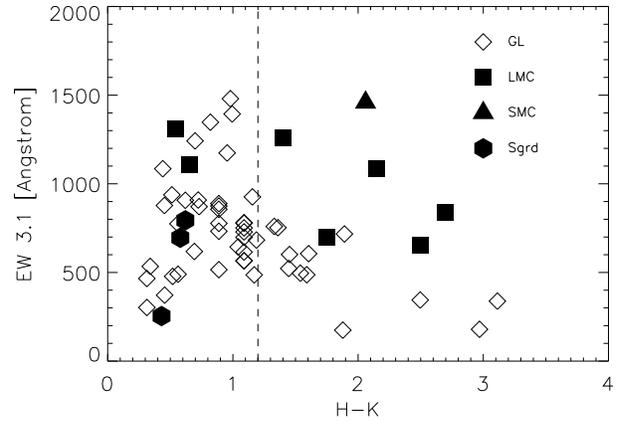}}
\caption{
The 
HCN and C$_2$H$_2$ equivalent widths of 
\Wo.
}
\label{Fig-eq31}
\end{figure}
%_________________________________________________________________

We measured the EWs of the three main molecular bands of carbon-rich
stars.  
The parameters used to measure the EW are summarized 
in
Table\,\ref{table-def}.  The EWs are plotted against infrared colour
in 
Fig.\,\ref{Fig-eq35}--\,\ref{Fig-eq31}, where the infrared colour
$H-K$ is 
used as a measure for the effective temperature.  The
circumstellar excess 
also affects this colour for redder carbon-rich
stars.

For comparison with 
Galactic stars, we reduced  ISO/SWS spectra of 
carbon stars in the solar 
neighbourhood, and also used the reduced ISO/SWS
spectra of \citet{Sloan03}. The 
infrared colours for the Galactic stars
were taken from the literature 
(Table\,\ref{table-carbon}).

As seen in the spectra, the 3.5\,$\mu$m EW is almost negligible in
extra-galactic AGB stars, in contrast to the nearby AGB stars which
have EWs up to 12\,\AA.  Only a few LMC stars show \Wf\, larger than
5\,\AA, but in those LMC stars that show the band,  
 this molecular band is as strong as in the
Galactic stars.  There is a clear tendency for the EW to decrease with
increasing H$-$K,  but there is also a group of
stars at blue H$-$K 
colour with very weak HCN bands.

T\,Dra, V\,CrB and V460 Cyg were observed 
with ISO/SWS at several phases.
We find evidence for a large  time variation 
in the HCN  \Wf; e.g.,
V\,CrB shows variation from 2--10\,\AA.

The 3.8\,$\mu$m 
EW, which probably measures the C$_2$H$_2$ abundance, is high
in red carbon 
stars ($H-K>1.2$), as shown in Fig. \,\ref{Fig-eq38}.  This
holds true for 
both the LMC stars and the single SMC star. The SgrD stars do
not show the 
band but this is consistent with their blue colour.
Among the Galactic 
stars, only a few objects show a 3.8\,$\mu$m
C$_2$H$_2$ band.

The 
3.1\,$\mu$m band is a mix of HCN and C$_2$H$_2$ bands.  \Wo\, is
large in 
the spectra of the extra-galactic, red carbon stars.  The
weakness of the 
3.5\,$\mu$m band in extra-galactic AGB stars suggests
a lower contribution 
from HCN in extra-galactic stars than in galactic
stars: the 3.1\,$\mu$m 
band of the extra-galactic stars may have a
larger contribution of 
C$_2$H$_2$ than Galactic stars.

Both \Wf\ and \We\ decrease towards redder 
infrared colours; \Wo\, increases
until $H-K$=1.2, and then \Wo\ begins to decreases. 
This colour corresponds roughly to
the tip of the AGB colour sequence: 
redder stars are affected by circumstellar
extinction and/or dust emission.  
The decline towards red colours suggests
that circumstellar emission fills 
in the absorption features.

\section{ Discussion}

\subsection{ SiO 
abundance in extra-galactic oxygen-rich AGB stars}

The SiO bonding energy 
is one of the highest (after CO) amongst the
major photospheric molecules. 
It is therefore not surprising that SiO
bands are commonly detected in 
late-type stars in our Galaxy,
including post-AGB stars \citep{Kaeufl92} and 
S-type stars
\citep{Wallace02}.  However, we only found one possible 
detection of
SiO bands among 6 LMC oxygen-rich stars (three of them are 
M-type
supergiants).  This difference needs to be explained.

The LMC 
oxygen-rich stars are very luminous \citep{vanLoon00}, and
three of our 
targets are supergiants.  Possibly they are more luminous
than the Galactic 
comparison stars. However, the luminosity class does
not affect the EW of 
the SiO bands, as shown by \citet{Wallace02} and
\citet{Heras02}.

The abundance of SiO is determined by the least abundant
component which will generally be silicon.
As an $\alpha$-element silicon is not synthesised in AGB stars
and its abundance is determined by the composition of the
progenitor star at its formation.
The major 
effect on SiO abundance should be the Si elemental
abundance.  The SiO 
abundance is limited by the least abundant
component, Si.  Silicon is an 
$\alpha$-element, not
synthesized in AGB stars; the Si abundance 
reflects the one at the
time the progenitor star was formed. In 
supergiants, Si is
formed but only in the last stage of their evolution.  
The current Si
abundance in the LMC can be measured from H{\sc II} regions 
or from
hot stars. The solar abundance is log(Si/H)+12=7.6 
\citep{Anders89}.
LMC H{\sc II} regions show log(Si/H)+12$=7.7$ 
\citep{Russell92}, and
main sequence B-type stars have $6.7$ 
\citep{Garnett99}.  For
planetary nebulae \citet{Dopita97} found 
log(Si/H)+12=6.5--7.7: the
large difference may reflect the range in 
progenitor ages. Here, we
use log(Si/H)+12$=7.0$ for the LMC late-type 
stars, taken from an
average over the planetary nebulae 
(PNe).

\begin{figure}
\centering
%\resizebox{\hsize}{!}{\includegraphics*{w40_metal.eps}}
\resizebox{\hsize}{!}{\includegraphics*{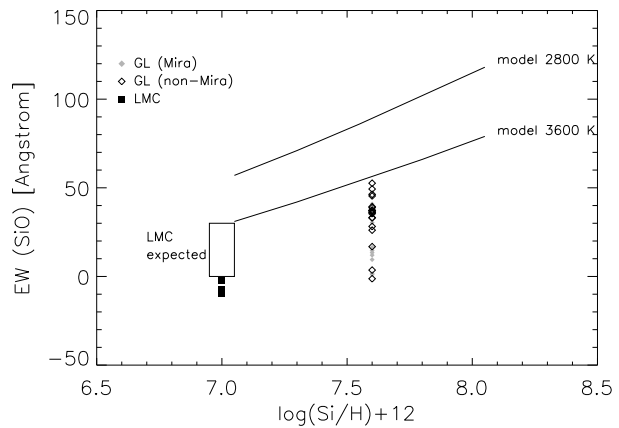}}
\caption{
The 
equivalent width of SiO bands as a function of metallicity.
The model is 
from \citet{Aringer97}, and Galactic data are 
from
\citet{Aringer99}
}
\label{Fig-eqmetal}
\end{figure}

\citet{Aringer97} calculated the EW of SiO bands in M-type 
giants
using a hydrostatic model.  \Ws\, is sensitive to both the 
effective
temperature and silicon abundance.  \citet{Aringer99} measured the 
SiO
EW of Galactic stars, finding values ranging from 0--50 
\AA.
Extrapolating this linearly to the LMC metallicity yields \Ws\, 
of
0--30 \AA\, (Fig. \,\ref{Fig-eqmetal}).  The low Si elemental
abundance 
could be the major cause of the lower \Ws\, observed in the
LMC.

Our 
measured \Ws\, limits for three stars are at the lower range of
the 
extrapolation from \citet{Aringer97, Aringer99}.  Galactic stars
show a 
spread in \Ws\,; for the LMC stars we can also expect a few
very low values. 
  The low upper limits can be interpreted as 
evidence that the Si abundance 
among LMC oxygen-rich stars is lower
than the assumed value of 7.0.  
However, two other possible reasons
should be considered.

First, for 
Galactic stars \citet{Aringer99} reported that observed
\Ws\, are lower than 
their model predictions, with Mira variables
showing weaker bands than 
semi-regular variables.  \citet{Tsuji94},
who observed M-giants and 
supergiants, suggest that the strong
pulsations extend the atmosphere above 
the photosphere.  Molecular
bands in this extended atmosphere may fill in 
their photospheric
counterparts.  In extreme cases, the SiO bands could be 
in emission
during the maximum phase of the variability \citep{Yamamura99}, 
when
the radius of the extended atmosphere is largest 
\citep{Matsuura02a}.
At least two of the LMC targets 
are Mira variables \citep{Whitelock03}.  We may have observed the 
stars
during the phase of the pulsation cycle when the SiO bands are
weak. 
For two stars, for instance, we know the approximate phase at the time of
observation: 
they were observed near maximum phase, when SiO bands could
be weak.

The 
second possible explanation is that dust emission suppresses the
SiO 
equivalent width. Among our Galactic sample, red stars (VY\,Sgr
and 
VY\,CMa) with H$-$K$> 0.7\,$mag do not show any trace of SiO
bands.  In 
our LMC sample, two stars have comparably high values in H$-$K
and could be affected by dust emission.
Therefore, we suggest that the dust
filling could decrease 
the \Ws.

We also investigated the effect of higher C/O abundance for LMC
stars, as dredge-up of newly produced carbon (3$^{\rm rd}$ dredge-up)
has a larger effect for low metallicity stars (see next section) and
would reduce the free oxygen after the initial formation of
CO. However, 
less oxygen has little or no effect on the SiO abundance
because the Si 
abundance still limits the SiO formation. Only for C/O
very near to unity 
(i.e. for S or SC stars) may some effect be expected
\citep{Zijlstra04}.

In 
conclusion, within our limited sample, we did not detect SiO bands at
all.  
The model predicts that low Si abundance results in the low SiO
abundance, 
and our observations do not conflict with this trend.  The
measured \Ws\, in 
the LMC is even lower than the expected range of \Ws.
Although several 
explanations, such as pulsations and dust filling,
can be given for the 
lower \Ws\, than expected in the LMC, the
issue is not settled. The solution 
awaits observation of a larger number of stars
with complete coverage in 
colour both in our Galaxy and in the 
LMC.

%_________________________________________________________________
%_________________________________________________________________

\subsection{ 
C$_{\boldmath{2}}$H$_2$, CH and carbon abundances of 
carbon-rich 
stars}

The 3.8\,$\mu$m C$_2$H$_2$ band is found in only a few ISO/SWS 
spectra
of Galactic carbon-rich stars: \object{V\,Cyg} and \object{HD\,192443}
\citep{Matsuura02b, Vandenbussche02}.  On the other hand, almost 
all
red LMC or SMC carbon stars show this band in their spectra.  
The
measured \We\, is clearly larger among red carbon stars in the 
LMC,
compared to the galactic stars. One SMC star measured in this work
also 
follows this trend.

The likely explanation for the stronger 3.8\,$\mu$m band is that the
LMC and SMC stars have a higher C/O ratio (Paper\,I). If all the
elemental abundances simply scale with the metallicity, the C$_2$H$_2$
abundance should be less because fewer carbon atoms are available. The
C/O ratio of the star, initially less than unity, increases during the
third dredge-up on the thermal-pulsing AGB.  A fixed amount of carbon
injected into the envelope has a larger effect for a low metallicity,
leading to a higher C/O ratio \citep{Lattanzio03, Mouhcine03}.
Additionally, the third dredge-up operates more efficiently at lower
metallicity \citep{Vassiliadis93}, adding larger quantities of carbon. 
Both effects predict a systematically higher C/O ratio in the LMC and the SMC. 
In stars with C/O$>$1 almost all the oxygen
is tied up in CO, and the excess carbon will be available for forming
carbon-bearing molecules.

A higher C/O ratio at low metallicity is 
supported by the fact that
the ratio of numbers of carbon-rich stars over 
oxygen-rich stars
($N_{\rm C}/N_{\rm O}$) increases with decreasing 
metallicity of the
host galaxy \citep{Groenewegen99}.  Further confirmation 
comes from
LMC and SMC planetary nebulae, which show a higher C/O 
ratio
compared to Galactic PNe \citep{Leisy96}. PNe show the final 
AGB
abundances after the last dredge-up event.

\citet{Cohen81} found that 
the $J-K$ colour is redder in LMC carbon
stars than in Galactic carbon 
stars. They suggested that the $J$-band
magnitude in LMC stars is suppressed 
by deeper C$_2$ absorption, and
also mention a higher C/O as the likely 
cause.  The stronger
C$_2$H$_2$ absorption is consistent with this.

\begin{figure*}
\centering
%\resizebox{\hsize}{!}{\includegraphics*{chemical_co.eps}}
\resizebox{\hsize}{!}{\includegraphics*{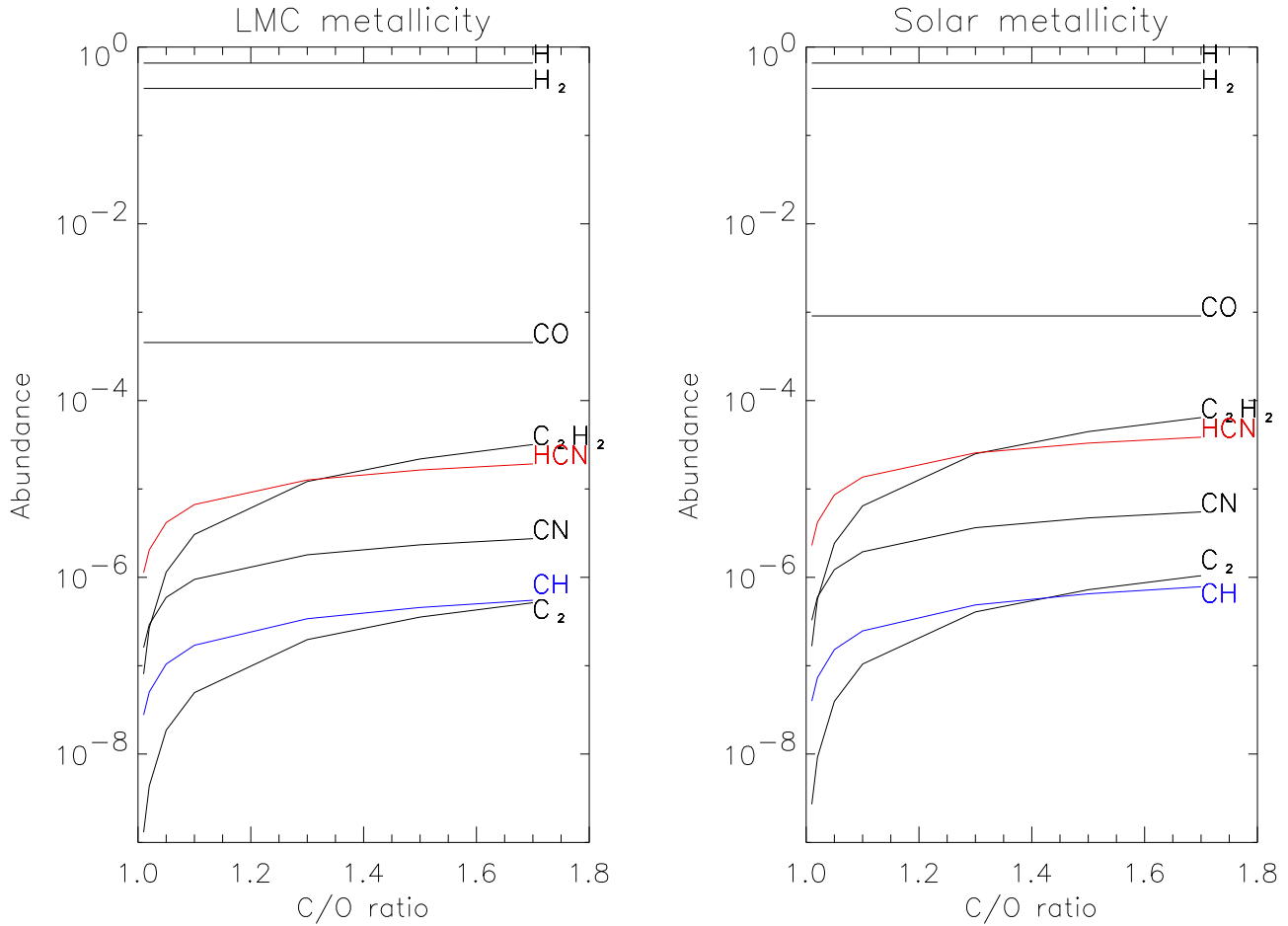}}
\caption{ 
Abundance of molecules calculated with the chemical
model. The abundance 
is normalized by H+H$_2$ pressure. The left side is for
[Z/H]=$-0.3$ 
(approximate LMC metallicity), and the right  for
[Z/H]=0.0 (solar 
metallicity). All elemental abundances are scaled to
the metallicity, except 
carbon, which is set relative to the oxygen
abundance using the C/O ratio.  
The temperature is 2500\,K for both panels.  
}
\label{Fig-chemical}
\end{figure*}

A 
chemical equilibrium model (Fig\,\ref{Fig-chemical}) shows 
increase of 
the C$_2$H$_2$ fractional abundance as function of the C/O
ratio.  Details of the model are given in \citet{Markwick00}.
Chemical equilibrium is a reasonable assumption 
for the stellar photosphere but could be questioned 
for the lower density atmosphere.  We assume a typical C/O
ratio for 
Galactic carbon stars of 1.2 \citep{Lambert86, Ohnaka00}.
To obtain a higher 
C$_2$H$_2$ fractional abundance for the LMC stars
requires C/O$\sim 1.4$, while 
the SMC carbon-rich stars should have an even
higher value.

The mean C/O 
ratio found in carbon-rich PNe in the LMC is 3.2 \citep[8
samples]{Leisy96} 
or 2.8 \citep[4 samples]{Dopita97}.  AGB stars may
experience several more 
third dredge-up events, giving a higher C/O
ratio in PNe than in AGB stars.  
However, we cannot exclude such a
high C/O ratio in the LMC AGB stars. The 
C$_2$H$_2$ abundance
increases only slowly for C/O$\,>$1.4 
(Fig\,\ref{Fig-chemical}); the
relative increase in the number of free 
carbon atoms (after CO
formation) is fastest for C/O close to unity.

\begin{figure*}
\centering
%\resizebox{\hsize}{!}{\includegraphics*{chemical_tmp.eps}}
\resizebox{\hsize}{!}{\includegraphics*{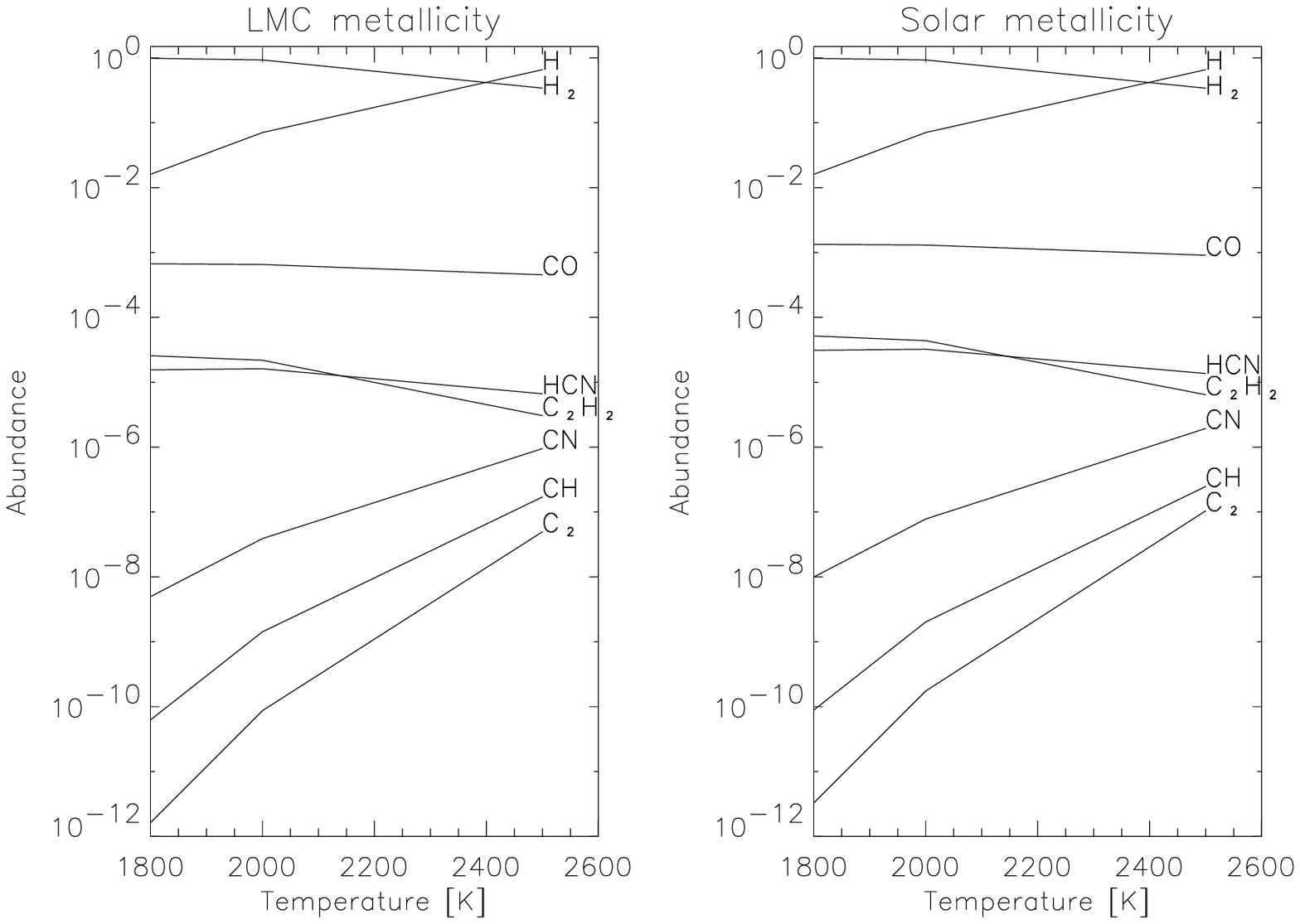}}
\caption{ 
Same as Fig.\ref{Fig-chemical}, but as a function of temperature. 
The C/O 
ratio is 1.1 for both 
panels.
}
\label{Fig-chemical2}
\end{figure*}

In 
Fig.\,\ref{Fig-eq38}, the EW of C$_2$H$_2$ increases sharply 
at
$H-K\approx0.5$, and declines above 1.2.  The initial increase is
caused 
by a temperature effect.  C$_2$H$_2$ is favoured at lower
temperatures: at 
higher temperatures more of the carbon is in
diatomic molecules. The 
temperature dependence of the various species
is shown in 
Fig\,\ref{Fig-chemical2}.  The decay at redder $H-K$
colour is explained by 
additional circumstellar emission.

We also detect CH bands in the 
infrared spectra, for carbon-rich stars
in the LMC, the SMC, and SgrD. The 
CH abundance depends on the C/O
ratio: \citet{Aoki98} found that SC-type 
stars, which have C/O ratio
close to one, show only weak CH lines, while 
N-type stars show clear CH
bands.  Fig.\,\ref{Fig-chemical} shows that
CH increases
more slowly with C/O ratio than does C$_2$H$_2$, as expected 
for its single
carbon atom. For a Galactic ratio of C/O$\,=1.2$, the CH 
abundances in the
LMC and the Galaxy may become comparable but one would not 
expect much
stronger bands in the LMC. The same argument holds for the SMC. 
We
have not made a comparison in line strengths of the CH bands because
they 
are blended with other molecular bands.  However,
the fact that these bands 
are seen in all stars including the Galactic
ones qualitatively supports the 
suggestion that the CH band strength
is relatively independent of 
metallicity.

 In conclusion, a high C/O ratio affects the spectra of 
extra-galactic
carbon-rich stars. The C/O ratio should be treated as an 
important
parameter for spectral 
classification.

%_________________________________________________________________
%_________________________________________________________________
%_________________________________________________________________
%_________________________________________________________________

\subsection{ HCN and nitrogen abundance of carbon-rich stars}

In contrast 
to the C$_2$H$_2$ and possibly CH, the HCN lines are generally
very weak in 
the extra-galactic stars.  \citet{LloydEvans80} finds that
optical carbon 
stars in the LMC show weak (optical) CN bands compared to
C$_2$. This argues 
against temperature as a factor, since C$_2$ and CN show
very similar 
temperature dependence (Fig\,\ref{Fig-chemical2}), as do (to a
lesser 
extent) C$_2$H$_2$ and HCN. This leaves the nitrogen abundance as the
most 
likely explanation.

Although the majority of extra-galactic carbon stars do 
not show the
3.5\,$\mu$m HCN feature, a few LMC stars have \Wf\,$>5\,$\AA
as high
as Galactic stars.  There may be two discrete extra-galactic 
groups:
HCN-rich and non-HCN. The C/N ratio should be different for 
these
two groups.

The C/N ratio varies strongly over the course of the 
star's evolution, mainly because of C production and CN cycling which 
transmutes C into N.  First dredge-up decreases carbon and increases 
nitrogen, and lowers the C/N ratio by a factor of 2 or more. The largest 
changes affect stars more massive than 2--3\,M$_\odot$.  Second 
dredge-up, which only occurs in stars with $M\gsim 4\,\rm M_\odot$, has a 
similar effect, and alters the C/N ratio by a factor of $\sim1.5$. 
Third dredge-up increases the carbon abundance by large factors but leaves 
N relatively unaffected. The most massive AGB stars may suffer burning at 
the bottom of the convective envelope, a process called hot bottom 
burning which again cycles C into N.  The cumulative changes in
the C/N ratios 
are shown in \citet{Boothroyd99}.

The effect of the third dredge-up scenario can explain two
distinct groups: low-metallicity stars without hot bottom burning
(HBB; \citet{Iben81}) will show low N, while stars with HBB show
enhanced N.  An example of an LMC star believed to have experienced HBB is
IRAS\,04496$-$6958, which is  a high-luminosity silicate carbon star 
\citep{Trams99a}.

The fact that 3 of our 8 LMC stars are HCN-rich suggests that the
lower mass limit for HBB cannot be much higher than the canonical
4\,$M_{\sun}$, since otherwise very few AGB stars would pass through
this phase.  In PNe, nitrogen and carbon elemental abundance ratios
vary greatly between individual objects \citep{Dopita97}, showing
the effect of the initial parameters on the enrichment history.

\subsection{ Grain formation and mass-loss at lower metallicity}

Our observations suggest that at low metallicities, carbon and
oxygen-rich stars behave differently. Carbon stars show higher C/O
ratios and more abundant carbon-bearing molecules (excluding CO) than
their Galactic counterparts.  Oxygen-rich stars show primarily low SiO
abundances, as expected from the low metallicity.  This difference can
be expected to affect their dust formation, since C$_2$H$_2$ and SiO are 
the respective starting points for build-up towards  either
carbonaceous or silicate dust grains.

For oxygen-rich stars, silicate grains require Si, O 
and iron and/or
magnesium.  The low SiO abundance caused by the low 
metallicity
suggests a lower silicate dust abundance. Iron or
magnesium may 
also limit the amount of silicate dust.

For carbon stars, 
\citet{Groenewegen95} shows that the mass of amorphous
carbon dust is about 10 
times more than that of silicon carbide (SiC).
Large molecules can build up 
from C$_2$H$_2$ \citep{Woods02} and can form a major component 
of carbonaceous dust, as shown by
the large mass of carbon dust.  The 
enhanced C/O ratio, giving a high
abundance of C$_2$H$_2$, can allow for 
carbonaceous dust to form at a
rate at least comparable to those of Galactic 
stars.  In addition, if
PAHs may be formed in AGB stars (but not excited), 
then more
C$_2$H$_2$ can also lead to enhanced PAH abundances.

The dust 
formation rate directly affects the mass-loss rate of the
star.  Above a 
metallicity of [Fe/H]$=-1.0$, the mass-loss process is
driven via radiation 
pressure on the dust -- the accelerated dust drags
the gas with it \citep{Bowen91}. 
At low metallicities the amount of dust formed will be very small and mass
loss will be very inefficient.

This all suggests that while oxygen-rich stars will show low mass-loss
rates due to the reduced SiO, carbon stars should be able to reach
mass-loss rates similar to Galactic stars,  which is consistent 
with our (limited) knowledge of LMC stars. Carbon stars in the LMC reach
mass-loss rates of a few times $10^{-5}\,\rm M_\odot\,yr^{-1}$
\citep{vanLoon00}, similar to what is found in the Galaxy.
Oxygen-rich stars in the LMC reach higher mass-loss rates than
Galactic stars \citep{vanLoon00}, but do so at a luminosity several
times higher than the LMC carbon stars.  Direct comparison with
Galactic stars is hampered by uncertain distances to Galactic
stars and by the fact that the LMC measurements were made for stars
known to be surrounded by dust, so that the sample is already
biased towards high mass-loss rates.

If oxygen-rich stars show 
reduced mass-loss but carbon stars do not,
this will affect the composition 
of the dust which subsequently enters the
ISM.  Even if oxygen-rich stars 
can still reach the mass-loss
rates, but do so later in their 
evolution, this would still imply a
shift in composition since the delay 
will allow more stars to reach
the carbon star phase.  The main sources of 
dust in a galaxy are
supernovae (and their progenitors) and AGB stars, but the relative
contribution of each is uncertain and will differ for silicate and
carbonaceous dust. Reduced mass-loss of oxygen-rich AGB stars will cause 
a relative shift towards carbonaceous grains.  The ISO
observations of 
\ion{H}{II} regions and molecular clouds show that PAH
bands are ubiquitous 
in the LMC and the SMC \citep{Reach00,
Vermeij02}, with a similar fraction 
to total dust as found in our
Galaxy;  the detailed composition of the PAHs 
may vary with
more open PAHs found in the SMC.  The fractional dust 
abundance in the
SMC is 30 times lower than that of our Galaxy, but although this holds
for both small and large grains \citep{Bot04}, it is not known whether the
fractions of carbonaceous and silicate dust differ. Our detection of a
PAH band from an SMC post-AGB star shows that PAHs can form in a
low-metallicity AGB wind.

\section{ Conclusion}
We report the strength of infrared molecular bands in
AGB stars in the LMC, SMC 
and SgrD galaxies. The metallicities
range from $-0.3$ to $-0.7$. Carbon stars 
are observed in all three
galaxies.  The LMC and the SMC contain luminous 
and high mass-loss
stars that are likely to be young, intermediate-mass 
AGB stars. The
SgrD stars are older with a likely age around 5\,Gyr and are 
optical carbon stars without evidence of high mass loss.

Our main conclusions are:

\begin{itemize}

\item The oxygen-rich stars in the sample show weak or no SiO
bands. We have good spectra for 3 stars at medium resolution, a lower
quality spectrum for one star, and two further spectra at low
resolution, all in the LMC.  At low metallicity, weaker SiO bands are
expected, and our measurements do not conflict with the metallicity
effects on SiO bands.  However, in our limited sample, factors 
such as dust or pulsation could hamper the correct measurements of SiO bands 
in the LMC.

\item The carbon-rich stars in the LMC and the SMC show very strong
C$_2$H$_2$ bands, when compared to Galactic stars, which implies a very high
abundance of this molecule is implied. The likely explanation is a high 
C/O ratio of low-metallicity carbon stars, leading to a larger amount of 
free carbon. The high C/O ratio is supported by data for planetary nebulae. 
Chemical equilibrium abundances are presented for a range of C/O ratios, 
showing that high C$_2$H$_2$ can be reproduced with C/O$\gsim 1.4$.

\item 
The HCN band is very weak compared to Galactic stars, except for a few LMC 
stars that show strong HCN. This could imply a difference in initial 
mass. We show that a low abundance of N-bearing molecules is expected for 
low-metallicity stars that do not experience hot bottom burning.  The two 
populations are likely to be separated in mass, since only stars with 
progenitor masses $ \gsim 4\,\rm M_\odot $ are likely to experience hot 
bottom burning.

\item CH bands are seen in stars from all galaxies, including our
own. The chemical equilibrium calculations show that this diatomic
molecule is less affected by the C/O ratio, so that similar abundances
in all the observed systems can be expected. The lines are blended 
with other molecular bands, and agreement with the models is
only qualitative.

\item We detect a PAH 3.3$\mu$m band in a post-AGB star in the SMC.
This may be the first detection of PAHs in such a star. 
PAH formation proceeds even in low-metallicity evolved stars.

\item The difference in abundance behaviour between carbon and oxygen
stars is expected to affect their dust formation and mass-loss rates.
In oxygen-rich stars, silicate dust forms, which should be directly
affected by the possible low abundance of SiO.  Carbon stars form
carbonaceous and hydrocarbon dust from molecules that we find to have
a high abundance. The mass-loss from AGB stars is driven by radiation
pressure on the dust, which leads us to expect that low-metallicity carbon
stars will show mass-loss rates as high as Galactic stars, but
oxygen-rich stars will tend to show lower mass-loss rates at the same
luminosity.  The limited available data indicates that some oxygen-rich
stars exist in the LMC with similar mass-loss rates to carbon stars,
but they are several times more luminous than the carbon stars. 
A comprehensive survey of mass-loss rates from LMC AGB stars is not 
yet available, but it could confirm whether less luminous
oxygen-rich stars in the LMC indeed show low mass-loss rates.

\item 
We predict that dust input from AGB stars into the ISM contains a higher 
fraction of carbonaceous grains than is the case for our Galaxy. Whether 
this affects the ISM dust depends on the relative contribution from AGB 
stars compared to other dust sources, primarily supernovae. PAH emission
is ubiquitous in the SMC, and our detection in the post-AGB star shows
that evolved stars are a potential source of PAHs.

\end{itemize}

%__________________________________________________________________

\begin{acknowledgements}
 
We would like to thanks Dr. Y. Jung for helping with our data reduction.
The ISO 
data archive and the 2MASS data archive provide useful data
for this study.  
Simbad data base is used for this research.  M.M. is
supported by a PPARC 
Rolling 
grant.
\end{acknowledgements}

\end{document}